\documentclass[aps,twocolumn,footinbib,nobalancelastpage,pra]{revtex4-2}
\usepackage{amsmath,amssymb}
\usepackage{graphicx}
\usepackage{amsmath}
\usepackage{slashed}
\usepackage{physics}
\usepackage{parskip}

\usepackage[table,xcdraw,dvipsnames]{xcolor}
\usepackage[hidelinks]{hyperref}
\usepackage{multirow}
\usepackage{braket}
\usepackage[caption=false]{subfig}

\begin{document}

\title{Demonstration of magic state power of $D(S_3)$ anyons with two qudits}

\author{Lucy Byles}
\email{pylsb@leeds.ac.uk}
\affiliation {\it School of Physics and Astronomy, University of Leeds, Leeds, LS2 9JT, United Kingdom}

\author{Ewan Forbes} 
\affiliation {\it School of Physics and Astronomy, University of Leeds, Leeds, LS2 9JT, United Kingdom}

\author{Jiannis K. Pachos}
\affiliation {\it School of Physics and Astronomy, University of Leeds, Leeds, LS2 9JT, United Kingdom}

\date{\today}

\begin{abstract}  

We consider a lattice of $d=6$ qudits that supports $\mathbf{D}(\mathbf{S}_3)$ non-Abelian anyons. We present a method for implementing both braiding and fusion evolutions using only the operators that create and measure anyons, without requiring additional dynamical control. This provides a minimal protocol demonstrating that $\mathbf{D}(\mathbf{S}_3)$ anyons can generate magic states, thereby establishing their universality for quantum computation. Furthermore, we show that the entire scheme can be encoded in just two qudits, offering a compact blueprint that is inherently scalable and readily implementable in current quantum platforms.

\end{abstract}

\maketitle 

\section{Introduction}

Anyons introduced a surprising shift in the paradigm of quantum computation. These quasiparticles with exotic statistics were first introduced as the means to perform fault-tolerant quantum evolutions \cite{leinaas1977theory, kitaev2003fault, nayak2008non}. As long as anyons remain intact, the quantum information encoded in them remains protected from destructive environmental perturbations. Soon anyonic quantum computation gave rise to a host of potential benefits for quantum information, such as the discovery of a new quantum algorithm for the efficient evaluation of Jones polynomials \cite{li2024photonic} 
and the proposal of new methods to combat thermal errors \cite{dennis2002topological, terhal2015quantum, roberts2020symmetry, bravyi2013quantum, yoshida2011feasibility, shen2022fracton, chamon2005quantum}. Central to these investigations is the consideration of anyonic models that are capable of universal quantum computation \cite{mochon2003anyons, cui2015universal, wootton2009universal, wootton2011universal}. Furthermore, anyonic systems can be directly translated into quantum error-correcting codes with high error thresholds and minimal resource requirements \cite{raussendorf2007topological, nielsen2004optical}.

The potential applications of anyons to quantum computation have sparked a race for their physical realisation. Fractional quantum Hall liquids \cite{moore1991nonabelions, das2005topologically, dolev2008observation, read2000paired} and topological superconductors \cite{ivanov2001non, xu2023digital, google2023non, mourik2012signatures} are the most promising platforms for the realisation of non-Abelian anyons capable of manipulating quantum information in a non-trivial way. Nevertheless, these platforms are highly complex, exhibiting a richness of physical effects \cite{yu2020non}. Moreover, anyons by their nature hide information rather well from the environment as well as from experimentalists. Hence, it is rather challenging to identify their properties in a physical system. To overcome some of these complexities, a significant effort has been dedicated in the physical simulation of quantum systems that give rise to certain properties of anyons \cite{aguado2008creation, xu2016simulating, liu2021topological, wootton2017demonstrating, stenger2021simulating}.
Such simulations allow for the effective isolation of the statistical properties of anyons from other physical effects, providing a platform to conclusively observe signatures of their non-Abelian behaviour~\cite{iqbal2024non}.

Here, we present an explicit demonstration of the non-Clifford action arising from the pairwise braiding of non-Abelian anyons in the framework of the $\mathbf{D}(\mathbf{S}_{3})$ quantum double model. In principle, the simulation of such braiding evolutions requires the physical transport of anyons \cite{aguado2008creation, lahtinen2010interacting, luo2011simulation}. Within a quantum simulation, however, such dynamical processes demand complex controlled evolutions capable of transporting anyons without performing unwanted measurement on their encoded information \cite{chen2025universal}. To bypass this complexity, we adopt an algebraic approach in which the ordered application of static lattice operations reproduces the effect of anyonic braiding in time rather than in space.
In this sense, our method offers a natural translation of Bonderson et al.’s `measurement-only' paradigm for topological quantum computation \cite{bonderson2008measurement, bonderson2009measurement} into the well-defined framework of Kitaev’s quantum double model. 

Concretely, by employing controlled sequences of localised ribbon and projection operators to a lattice of $d=6$ qudits, we simulate the action of the fundamental exchange and fusion recombination matrices, $R$ and $F$, on the topologically encoded fusion space of non-Abelian $G$ anyons. These matrices enable the reconstruction of the braiding operators acting on the fusion space of two anyon pairs. 
Our aim is to present the manipulations required to experimentally demonstrate that the $D(S_3)$ model can generate magic states through braiding operations \cite{bravyi2005universal, bravyi2012magic}. Notably, such magic states are a valuable resource in bridging between efficiently simulatable Clifford circuits and full universal quantum computation \cite{howard2017application}. 

Further analysis of the operator structure reveals that all non-trivial action of the ribbon and projection operators for the $\{A,B,G\}$ sub-model is encoded in their action on just two $d=6$ qudits. In this way, only a minimal system of two qudits is required to reconstruct the elements of the fundamental $R$ and $F$ matrices. This dense encoding enables a significant reduction in required experimental resources, rendering this scheme accessible to various quantum technology platforms, ranging from photonic systems \cite{lu2009demonstrating, pachos2009revealing, goel2023unveiling} to Rydberg atoms \cite{weimer2010rydberg, benhemou2023universality}. This minimal scheme also offers direct generalization to systems with a greater number of qudits, where the topological invariance of the evolutions under continuous deformations can be demonstrated.
Furthermore, the compact topological encoding of information in the fusion space of four non-Abelian $G$ anyons presented here offers a scalable blueprint for extending the scheme to multiple anyons, enabling the implementation of logical entangling gates and prototype topological quantum algorithms~\cite{hormozi2007topological, field2018introduction, aharonov2006polynomial}.

This paper is organised as following. In Section II we review the main elements of the $\mathbf{D}(\mathbf{S}_{3})$ quantum double model and present a specific subset of non-Abelian anyons it supports that can give rise to magic states. A minimal set of steps that determine the $R$ matrix are given in Section III, while a simple method that determines the fusion recombination $F$ matrix is given in Section IV. The obtained matrices are shown to explicitly demonstrate the non-Clifford action of the braiding of $G$ anyons, highlighting their importance as a resource for universal quantum computation.
In Section V, a dense encoding scheme is presented, whereby only two $d=6$ qudits are required to obtain the $R$ and $F$ matrices of the $G$ non-Abelian anyons. Finally, Section VI provides concluding remarks and outlook.

\section{Magic state preparation with $\mathbf{D}(\mathbf{S}_{3})$ anyons}
\label{sec:double_mod}


Quantum double models were first introduced by Alexei Kitaev in his seminal 2003 paper~\cite{kitaev2003fault} as a framework for encoding and manipulating quantum information in a way that intrinsically suppresses errors, thus paving the way toward scalable, fault-tolerant quantum computation. These models realise generalised quantum error-correcting codes as condensed matter systems, where a gapped Hamiltonian energetically protects the encoded logical information from erroneous perturbations. 

The quantum double models are defined on a lattice of qudits with spin levels labelled by the elements of a chosen finite group, $\mathbf{G}$. The group $\mathbf{G}$ defines the algebraic structure of the quantum double $\mathbf{D}(\mathbf{G})$, dictating the anyons present and their associated fusion and braiding relations. 
Abelian groups, such as the toric code $(\mathbf{G}=\mathbf{Z}_{2})$ for example, give rise to Abelian anyons only, whose braiding at most induces phase factors on the global wavefunction \cite{pachos2012introduction}. When combined with logical encodings and projective measurements of fusion outcomes, these processes act to map Pauli operators to Pauli operators, thereby realising the Clifford group \cite{dennis2002topological,bombin2006topological}. Formally, the $n$-qubit Clifford group
\begin{equation}
    \mathcal{C}_{n}=\{U | UPU^{-1}\in \mathcal{P}_{n} \forall P\in\mathcal{P}_{n}\},
\end{equation}
is the set of unitary operations that map the $n$-fold Pauli group $\mathcal{P}_{n}=\{\pm1,\pm i\}\times\{\mathbf{1}_{2},\sigma^{x},\sigma^{y},\sigma^{z}\}^{\otimes n}$ to itself \cite{grier2022classification}.
By the Gottesman–Knill theorem \cite{gottesman1998heisenberg, aaronson2004improved}, all such Clifford operations can be efficiently simulated on a classical computer, rendering such Abelian models insufficient for the implementation of a universal quantum computer.

Extension beyond classical simulability therefore requires the consideration of quantum double models based on non-Abelian groups. 
For this purpose, we consider the quantum double based on $\mathbf{S}_{3}$, the smallest non-Abelian group. As remarked by Lo {\em et al.} in their review \cite{lo2025universal}, $\mathbf{D}(\mathbf{S}_{3})$ lies at the `sweet spot' for the realisation of non-Abelian topological order, being both solvable and non-nilpotent \cite{dummit2004abstract}. Indeed, it has been shown that the non-Abelian anyons of this model can perform universal quantum computation through braiding and fusion alone \cite{mochon2004anyon, cui2015universal}. Furthermore, as $\mathbf{S}_{3}$ is the smallest non-Abelian group, the lattice needed to realise such a model has a relatively small local Hilbert space, rendering it more readily accessible to current small-scale quantum platforms \cite{goel2023unveiling}.

In recent years, this model has also gained renewed attention with the demonstration that the braiding of certain non-Abelian anyons within this model enables the generation of `non-stabilizer' or `magic' states \cite{laubscher2019universal}. These states are a valuable resource for universal quantum computation, extending the computational power of Clifford gates to allow access to a dense subset of the unitary group on the qubit Hilbert space \cite{bravyi2005universal}. In this way, classically simulatable stabilizer circuits may be promoted to full quantum universality through the `injection' of these special states \cite{knill2005quantum, howard2014contextuality}. 
The role of magic as a computational resource may be formalised through the
stabilizer Rényi entropy, $M_{\alpha}$ \cite{howard2017application, tarabunga2023many}. This provides a quantitative measure of the magic content of a quantum state, capturing the extent to which it deviates from a stabilizer state. For an $n$-qubit state $\ket{\psi}$, it is defined in terms of the squared overlaps with elements of the $n$-qubit Pauli group $\mathcal{P}_{n}$ as
\begin{equation}
    M_{\alpha}(\ket{\psi}) = \frac{1}{1-\alpha}\log{\left(\sum_{p\in\mathcal{P}_{n}} \frac{|\bra{\psi}p\ket{\psi}|^{2\alpha}}{2^{n}} \right)},
    \label{eq:SRE}
\end{equation}
where $\alpha>0$ is the Renyi index. $M_{\alpha}=0$ indicates a classically simulatable state, while larger values of $M_{\alpha}$ indicate greater deviation from the Clifford-stabilizer set, and hence higher `magic'.
Typically, we refer to the second-order stabilizer Renyi entropy, which for a single qubit state, is bounded as $0\leq M_{2}\leq \log{2}$.

Within quantum circuit architectures, magic states are typically prepared using magic state distillation protocols \cite{campbell2012magic, anwar2012qutrit, kim2024magic}. The probabilistic nature of these schemes however, generally incurs significant overhead in both qubit number and circuit depth, posing a significant challenge for scalable implementation \cite{liu2023magic, campbell2009structure}. Recent proposals for realizing robust magic state generation through the braiding of non-Abelian anyons therefore offer an appealing route toward inherently fault-tolerant universal quantum computation \cite{bombin2006topological, huang2025generating}. In particular, the work of Laubscher {\em et al.} \cite{laubscher2019universal} demonstrates the potential for achieving such robust magic state preparation within the $\mathbf{D}(\mathbf{S}_{3})$ quantum double model. The aim of the present work is to translate this approach into explicit lattice operations, enabling direct implementation within a quantum simulator. In the following, we begin with a brief overview of the relevant features of the $\mathbf{D}(\mathbf{S}_{3})$ quantum double model (expanded upon in Appendix \ref{sec:S3}), including a demonstration of magic state generation through non-Abelian braiding.

\subsection{The $\mathbf{D}(\mathbf{S}_{3})$ Quantum Double Model}
\label{sec:DS3}

The $\mathbf{D}(\mathbf{S}_{3})$ quantum double model describes eight distinct types of anyon, labelled $\{A, B, C, D, E, F, G,H\}$, corresponding to each of the eight irreducible representations of the quantum double of $\mathbf{S}_{3}$ \cite{beigi2011quantum, komar2017anyons}. Instead of considering the complete set of eight anyons, it is instructive to restrict to a smaller sub-group of this model that is closed under fusion. In this way, the non-Abelian properties can be demonstrated, while reducing the complexity of theoretical and experimental treatment. Here, we will consider the sub-group $\{A,B,G\}$ where $A$ is the vacuum, $B$ is an Abelian anyon and $G$ is a non-Abelian anyon. Their fusion relations are given by
\begin{equation}
    \begin{gathered}
        A\times A = B\times B=A, \\
        G\times B=G, \\
        G\times G=A+B+G.
    \end{gathered}
    \label{eq:fusionrel}
\end{equation}
The fusion multiplicity of the $G$ anyon signals its non-Abelian nature such that the fusion space of a pair of $G$ anyons forms a multi-dimensional Hilbert space spanned by a set of anyonic basis states $\ket{G,G\rightarrow i}$ where $G,G\rightarrow i$ indicates the fusion of two $G$ anyons to outcome $i=A,B,G$.
Notably, this fusion space is a non-local degree of freedom: it cannot be accessed or manipulated through local operations or classical communication (LOCC) on individual anyons. The encoding of logical states within this fusion space, therefore provides a means of encoding quantum information non-locally such that it is intrinsically protected from local environmental perturbations. 
The controlled manipulation of the fusion space, and therefore the encoded quantum information, is achieved using non-local topological operations such as anyonic braiding and fusion that are encoded in the so called $R$ and $F$ matrices \cite{nayak2008non}. 




\begin{figure}[!t]
    \centering    \includegraphics[width=0.45\textwidth]{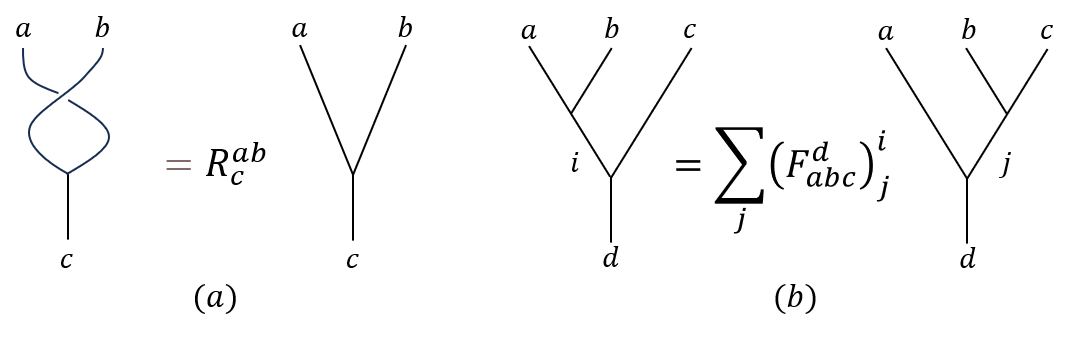}
    \caption{The $R$ and $F$ operations that characterise anyonic models. (a) The $R$ braid describes the phase, $R^{ab}_c$, gained by exchanging two anyons, $a$ and $b$ when they have a fixed fusion outcome $c$. (b) The $(F^d_{abc})^i_j$ matrix elements give the relation between the states of three fusing anyons $a$, $b$, and $c$ with fixed fusion outcome $d$ when the order of fusion is changed.}
    \label{fig:R_and_F}
\end{figure}


The $R$ matrix dictates the change in phase gained when two anyons undergo a counter-clockwise exchange, given as
\begin{equation}
R_{\circlearrowleft}\ket{a,b\rightarrow c} = R^{ab}_{c}\ket{b,a\rightarrow c},
    \label{eq:Rabc}
\end{equation}
for two anyons $a$ and $b$ fusing to an anyon $c$, where $R_{\circlearrowleft}$ corresponds to the anti-clockwise exchange of $a$ and $b$ anyons, as shown in Figure \ref{fig:R_and_F}(a). For the non-Abelian fusion of two G anyons, the fusion multiplicity generates a $3\times3$ diagonal unitary matrix $R^{GG}$ with entries $(R^{GG} )_{i}^{i}\equiv R_{i}^{GG}$, the form of which may be analytically determined using a set of self-consistent relations known as the pentagon and hexagon relations \cite{pachos2012introduction}. The matrix $R^{GG}$ is found to take the form
\begin{equation}
    R^{GG} = \begin{pmatrix}
        \omega & 0 & 0 \\
        0 & -\omega & 0 \\
        0 & 0 & \bar{\omega}
    \end{pmatrix}, \label{eq:R_matrix}
\end{equation}
with $\omega=e^{\frac{2\pi i}{3}}$ and $\bar{\omega}=\omega^{*}$.


The fusion matrix, $F$, describes the mapping between two states of three anyons $a, b$ and $c$ fusing to a final anyon, $d$, with the order of fusion changed as in Figure \ref{fig:R_and_F}(b). The action of this mapping may be described in the anyonic basis as
\begin{equation}
    \ket{(a,b),c\rightarrow i,c\rightarrow d} = \sum_{j} (F^{d}_{abc})^{i}_{j} \ket{a,(b,c)\rightarrow a,j\rightarrow d}.
    \label{eq:F}
\end{equation}
For the fusion of three $G$ anyons to total fusion outcome $G$, the intermediary anyons $i$ and $j$ may be $A,B$ or $G$. $F_{GGG}^{G}$ is therefore a $3\times3$ matrix. As with the $R$ matrix, the forms of these matrices may be found analytically from a set of consistency relations defined by the group structure \cite{cui2015universal, Siehler_2003}. For the $G$ anyons of the $\mathbf{D}(\mathbf{S}_{3})$ quantum double model it is found that
\begin{equation}
    F^G_{GGG} = \frac{1}{2} \begin{pmatrix}
        1 & 1 & \sqrt{2} \\
        1 & 1 & -\sqrt{2} \\
        \sqrt{2} & -\sqrt{2} & 0
    \end{pmatrix}. 
    \label{eq:F_matrix}
\end{equation}

The $R$ and $F$ matrices completely encode the description of all unitary braiding evolutions associated with a given anyonic model \cite{tounsi2023systematic, simon2023topological}. In the following, we explicitly derive the braid group representation on the fusion space of four $G$ anyons. Our protocol uses minimal resources to reveal the non-Clifford nature of $G$-anyon braiding, thereby paving the way for experimental demonstrations of their utility as a resource for universal quantum computation on current quantum hardware.


\subsection{Magic state generation from anyonic braiding}

\begin{figure}[!t]
    \centering
    \includegraphics[width=0.47\textwidth]{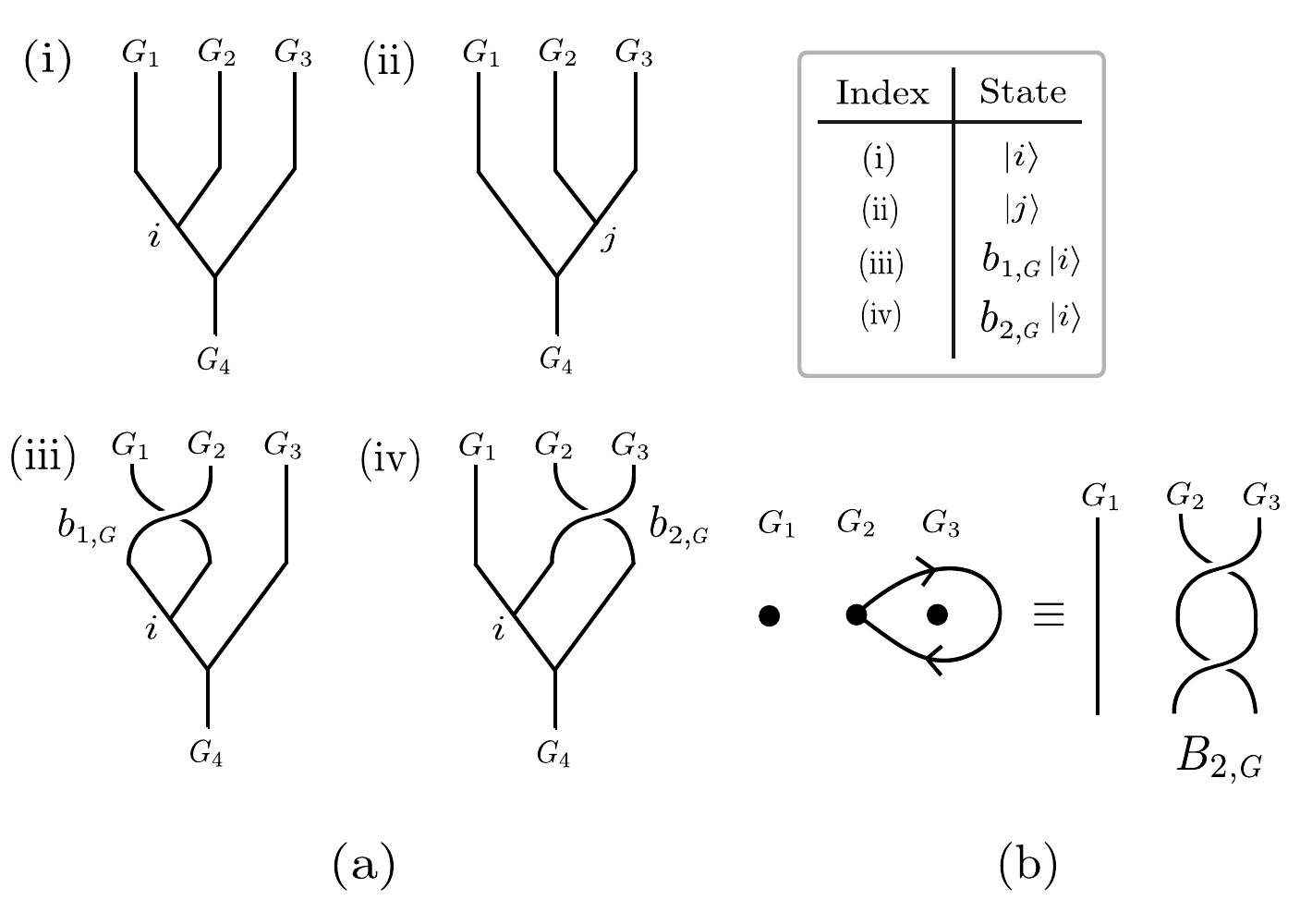}
    \caption{(a) The states (i) $\ket{i}$ and (ii) $\ket{j}$ belong to alternative bases for the fusion space of four $G$ anyons. The generators $b_{1,G}$ and $b_{2,G}$ for the three-strand braid group $\mathcal{B}_{3}$ act on the states $\ket{i}$ as shown in (iii) and (iv) respectively. (b) Repeated application of the generator $b_{2,G}$ generates a complete evolution of $G_{2}$ around $G_{3}$ as represented by the operation $B_{2,G}\equiv (b_{2,G})^{2}$.}
    \label{fig:b1b2}
\end{figure}

Consider now the fusion of three $G$ anyons $G_{1}, G_{2}$ and $G_{3}$ to fixed outcome $G_{4}$ as shown in Figure \ref{fig:b1b2}. With this fixed order of fusion the three-dimensional basis may be explicitly spanned by the states $\{\ket{i}\}\equiv\{\ket{(G_{1},G_{2}),G_{3}\rightarrow i,G_{3}\rightarrow G_{4}}\}$ for $i=A,B,G$. 
Alternatively, anyonic associativity means that this fusion of three $G$ anyons to a fixed outcome may equivalently correspond to the basis  $\ket{j}=\ket{G_{1},(G_{2},G_{3})\rightarrow G_{1},j\rightarrow G_{4}}$, with the two bases related by the matrix $F^{G}_{GGG}$ through $\ket{i} = \sum_{j} (F^{G}_{GGG})^{i}_{j} \ket{j}$, and $\ket{j} = \sum_{i} (F^{G \hspace{0.2cm}-1}_{GGG})^{j}_{i} \ket{i}$.

The topologically distinct paths of $G_{1}, G_{2}$ and $G_{3}$ prior to fusion form the 3-strand braid group $\mathcal{B}_{3}$. This group has two generators $b_{1,G}$ and $b_{2,G}$ braiding pairs $(G_{1},G_{2})$ and $(G_{2},G_{3})$ respectively as shown.
When acting on a defined fusion space, the action of these braiding operations may be expressed as a product of $R$ and $F$ matrices. Consider for example $b_{1,G}\ket{i}$ as shown in Figure \ref{fig:b1b2}. The anyons $G_{1}$ and $G_{2}$ share a direct fusion channel such that we have simply
\begin{equation}
    b_{1,G}\ket{i} = R^{GG}\ket{i}.
    \label{eq:b1G}
\end{equation}
In contrast, in the basis $\{\ket{i}\}$ the anyons $G_{2}$ and $G_{3}$ do not fuse directly and thus the matrix $F^{G}_{GGG}$ must be used to form a description of the operation $b_{2,G}$ \cite{simon2023topological}. Following the diagrammatic proof of Figure \ref{fig:b2proof} (details in Appendix \ref{sec:mbraid}), one finds
\begin{equation}
    b_{2,G}\ket{i} = F^{G \hspace{0.1cm} -1}_{GGG}R^{GG}F^{G}_{GGG}\ket{i}.
    \label{eq:b2G}
\end{equation}

\begin{figure}[!t]
    \centering
    \includegraphics[width=0.45\textwidth]{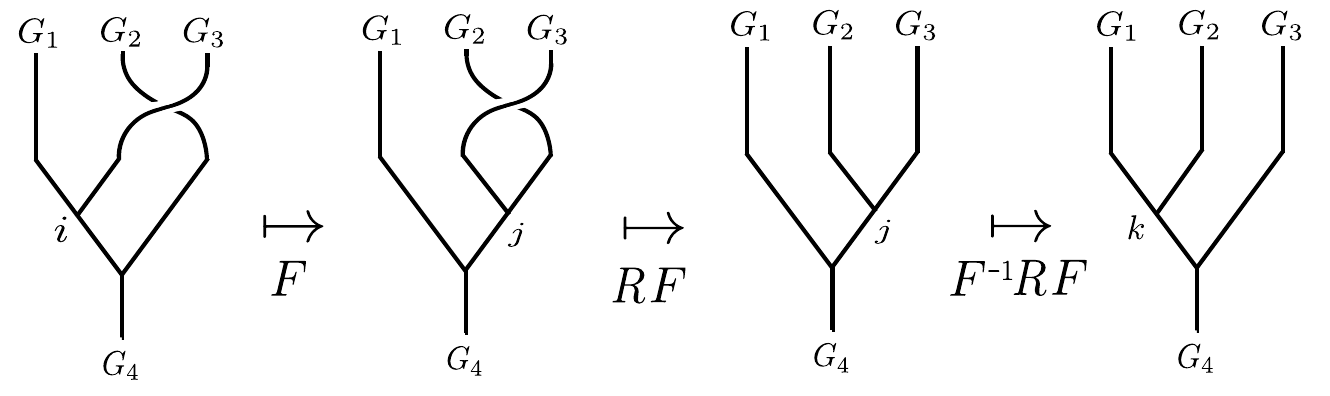}
    \caption{The action of the braiding operator $b_{2,G}$ on the basis state $\ket{i}$ can be understood through a series of $R$ and $F$ moves as shown.}
    \label{fig:b2proof}
\end{figure}

Direct substitution of $R^{GG}$ and $F^{G}_{GGG}$ thus yields exact expressions for the generators $b_{1,G}$ and $b_{2,G}$. 
As will be discussed further in Section \ref{sec:R} however, the single exchange $R$ is not directly represented as a physical operation within the formalism of the quantum double model. Instead, the observable that naturally appears from commutation relations of ribbon operators is the monodromy (full braiding) operator $R^{2}$ \cite{buerschaper2009mapping, bombin2008family}.
In this protocol we therefore consider the full braiding operators
\begin{align}
    & B_{1,G}\equiv b_{1,G}^{2}= (R^{GG})^{2}, \nonumber \\ 
    & B_{2,G}\equiv b_{2,G}^{2}=F^{G \hspace{0.1cm} -1}_{GGG}(R^{GG})^{2}F^{G}_{GGG},
    \label{eq:B2G}
\end{align} describing complete evolutions of $G_{1}$ around $G_{2}$ and $G_{2}$ around $G_{3}$ respectively (see Figure \ref{fig:b1b2}). 
Explicitly, $B_{2,G}$ has the following action on the anyonic basis $\{\ket{A},\ket{B},\ket{G}\}$
\begin{equation}
    B_{2,G} = \begin{pmatrix}
        \cos{\left(\frac{2\pi}{3}\right)} & -i\sin{\left(\frac{2\pi}{3}\right)} & 0\\
        -i\sin{\left(\frac{2\pi}{3}\right)} & \cos{\left(\frac{2\pi}{3}\right)} & 0 \\
        0 & 0 & \bar{\omega}
    \end{pmatrix},
\end{equation}
revealing a splitting in the qutrit subspace as this operation preserves the two-dimensional subspace $\text{span}(\ket{A},\ket{B})$ and its orthogonal complement $\text{span}(\ket{G})$. $B_{2,G}$ is thus a well-defined operation on a qubit encoded in terms of the reduced basis $\{\ket{A},\ket{B}\}$. Notably, inspection of this logical operation reveals that it does not belong to the single-qubit Clifford group $\mathcal{C}_{1}$.
Furthermore, creating two pairs of $G$ anyons from the vacuum and performing the braiding operation $B_{2,G}$ enables the preparation of the state
\begin{equation}
    \ket{\psi} = B_{2,G}\ket{A} = \cos{\left(\frac{2\pi}{3}\right)}\ket{A} -i\sin{\left(\frac{2\pi}{3}\right)}\ket{B},
\end{equation}
a magic state with $M_{2}(\ket{\psi})=\log{\left(\frac{16}{13}\right)}$. This non-stabilizer state constitutes a crucial resource for quantum computation, whereby the encoded non-Clifford action
enables the preparation of arbitrary quantum states using circuits composed solely of classically simulable Clifford operations \cite{laubscher2019universal}.

To illustrate the relative computational power of these $G$ anyons, we provide a brief comparison with an alternative non-Abelian subgroup of $\mathbf{D}(\mathbf{S}_{3})$, the charge subgroup $\{A,B,C\}$. 
As outlined in Table \ref{tab:ds3 anyons}, the non-Abelian chargeon $C$ has identical fusion rules to that of the $G$ anyon, such that a set of four $C$ anyons similarly encodes a logical qutrit. The form of the braiding matrix  
\begin{equation}
     R^{CC} = \begin{pmatrix}
        1 & 0 & 0 \\
        0 & -1 & 0 \\
        0 & 0 & 1
    \end{pmatrix},
\end{equation}
acting on the basis $\{\ket{A},\ket{B},\ket{C}\}$ however, yields $(R^{CC})^{2}=\mathbf{1}_{3}$. As a result the braiding operators $B_{1,C}=(R^{CC})^{2}=\mathbf{1}_{3}$ and $B_{2,C}=F^{C \hspace{0.1cm} -1}_{CCC}(R^{CC})^{2}F^{C}_{CCC}=\mathbf{1}_{3}$ act trivially on this fusion space, rendering this alternative subgroup unable to manifest non-trivial braiding statistics within the quantum double model.

\subsection{Anyonic manipulation on a lattice}
\label{sec:DG}

\begin{figure}[!t]
    \centering
    \includegraphics[width=0.45\textwidth]{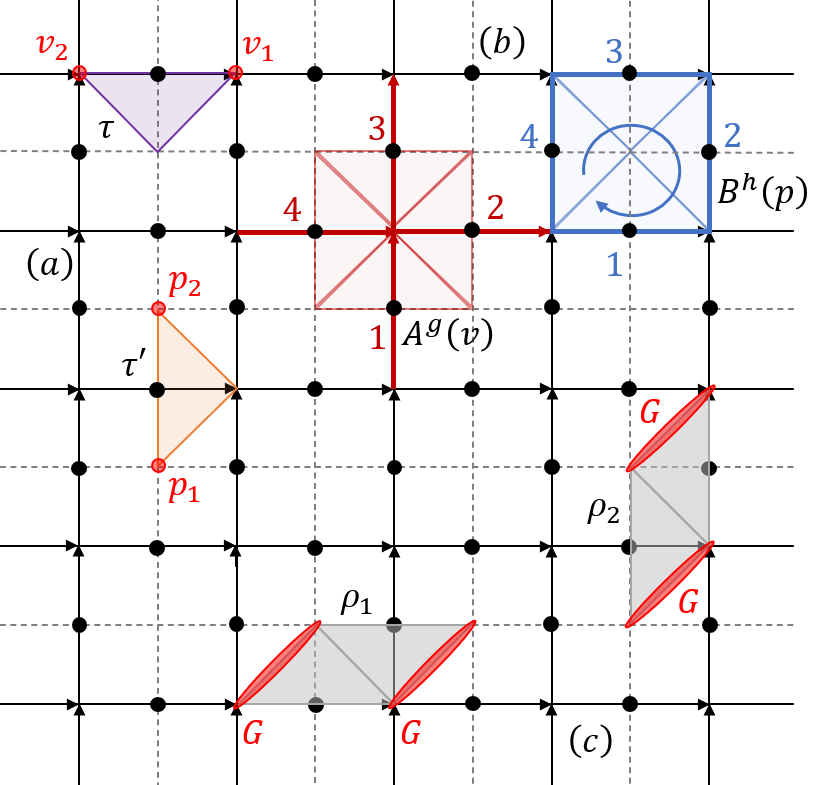}
    \caption{The lattice construction of the quantum double model $\mathbf{D}(\mathbf{S}_{3})$. The solid lines denote the direct lattice with underlying orientation as indicated by arrows. The dual lattice is shown with dotted lines. At the top of (a), an example of a direct triangle is shown as outlined in equation \eqref{eq:T}. The direct triangle, $\tau$, creates excitations on vertices $v_{1}$ and $v_{2}$ as shown. Below, is a dual triangle as defined in equation \eqref{eq:L}. The dual triangle, $\tau^{\prime}$, creates a pair of excitations on plaquettes $p_{1}$ and $p_{2}$.  In (b) closed loops of these dual (direct) triangles form the vertex (plaquette) operators. The orientation of each of the component operators depends on the lattice enumeration and ensures that all $A^{g}(v)$ and $B^{h}(p)$ mutually commute. (c) The two $G$ anyonic ribbons $\rho_{1}$ and $\rho_{2}$ that will be used throughout the paper. The dyons created by the associated ribbon operators $F^{G}_{\rho_{1}}$ and $F^{G}_{\rho_{2}}$ are situated along the sites formed of the vertices and plaquettes at the endpoints of each ribbon as highlighted.}
    \label{fig:Lattice}
\end{figure}


We now present the lattice model that gives rise to the quantum double $\mathbf{D}(\mathbf{S}_{3})$ presented above. Consider a two-dimensional square lattice with orientation as shown in Figure \ref{fig:Lattice}. On each link of the lattice a Hilbert space, $\mathbf{H}$, described by $d=6$ qudits ($\text{dim}(\mathbf{H})=6$) is placed with orthonormal basis indexed by the group elements of $\mathbf{S}_{3}$, $\{\ket{g}:g\in\mathbf{S}_{3}\}$

The dynamics of our model are given by the Kitaev Hamiltonian \cite{kitaev2003fault}
\begin{equation}
    \mathcal{H} = -\sum_{v} A(v) -\sum_{p} B(p),
    \label{eq:H}
\end{equation}
for vertex and plaquette operators $A(v)$ and $B(p)$ respectively as shown in Figure \ref{fig:Lattice}(b), defined on a lattice with open direct lattice boundaries. Hamiltonian terms outside of these boundary lines are set to zero \cite{cong2017universal}. This model produces a unique ground state $\ket{\zeta}$, corresponding to the anyonic vacuum, in which 
\begin{equation}
    A(v)\ket{\zeta}=\ket{\zeta}, \hspace{0.2cm} B(p)\ket{\zeta}=\ket{\zeta},
    \label{eq:gscond}
\end{equation} 
for all vertices, $v$, and plaquettes, $p$. Excitations of this model are particle-like and are indicated by violations of the conditions \eqref{eq:gscond}. Fluxons (chargeons) are anyonic excitations with non-trivial flux (charge) positioned on the plaquette (vertex) of the lattice for which $B(p)\ket{\zeta}\neq\ket{\zeta}$ ($A(v)\ket{\zeta}\neq\ket{\zeta}$). This model also admits the description of dyons$-$ composite particles defined on a neighbouring vertex and plaquette. 

The anyons in our reduced model $\{A,B,G\}$ may be coherently created and manipulated with the application of so-called ribbon operators to this anyonic ground state. As Hamiltonian~\eqref{eq:H} is a sum of commuting operators, only topological characteristics of the anyonic ribbon operators affect the logical encoding, with small perturbations leaving it invariant. The $A$ and $B$ anyons are both Abelian with trivial flux, and can therefore be created and moved with string operators corresponding to paths on the direct lattice. 
As $A$ is the vacuum particle, the operator $F^A$ acts trivially as the identity matrix. Both $A$ and $B$ anyons may be transported around the lattice similarly to $e$ and $m$ particles in the toric code by applying these single-qudit ribbon operators to strings of qudits.
In contrast, the $G$ anyons are dyons that have both non-trivial flux and charge. They therefore require the implementation of ribbon operators composed of dual and direct triangles for their manipulation \cite{bombin2008family} (see also Appendix \ref{sec:DGribbons}). A simple ribbon operator capable of producing a pair of $G$ anyons can be composed of one direct and one dual triangle, $\rho=\tau_{\text{direct}}\tau_{\text{dual}}$. In the following we will consider two distinct two-qudit ribbon operators, $F^{G}_{\rho_{1}}$ and $F^{G}_{\rho_{2}}$, as illustrated in Figure \ref{fig:Lattice}(c). We note that both of these operators are Hermitian but not unitary (see Appendix \ref{sec:LatticeRep} for explicit forms of these operators). Nevertheless, when restricted to act on the eigenstates of \eqref{eq:H}, they return normalised states \cite{luo2011simulation}. These ribbons are sufficient to demonstrate the fusion and braiding properties of the $G$ anyons of $\mathbf{D}(\mathbf{S}_{3})$, given in \eqref{eq:R_matrix} and \eqref{eq:F_matrix}, as we shall see in the next sections.

Finally, we will introduce a set of local measurement operators that distinguish between each anyon type. This is particularly useful when we want to determine the fusion outcome of two anyons while constructing anyonic basis states such as $\ket{G,G\rightarrow i}$, with $i=A, B, G$. Locality dictates that when two anyons fuse together to produce some overall particle type then both fusing particles must be measured jointly \cite{simon2023topological}. The initial anyons may be spatially separated, in which case the fusion outcome measurement must be performed around a loop that surrounds both anyons under fusion. For the operations considered here we ensure that the pair of particles undergoing fusion share a common vertex. In this way each outcome $A,B,G$ can be measured by the respective four-body vertex projection operator $A^{A/B/G}(v)$ as in \cite{komar2017anyons}, which project onto the chosen anyon by measuring the charge. These vertex projection operators form a set of orthogonal projective measurements as indicated by the identities
\begin{equation}
    \begin{split}
        \sum_{i\in\{A,B,G\}} A^{i}(v) = \mathbf{1}, \hspace{1.7cm}\\
        A^{i}(v)A^{j}(v) = \delta_{i,j} A^{i}(v), \hspace{0.4cm} i,j=A,B,G,\\
    \end{split}
    \label{eq:projorth}
\end{equation}
and are thus sufficient in uniquely distinguishing between the possible anyonic charges that lay on a chosen vertex $v$. 
For example for the ground state anyonic vacuum, $\ket{\zeta}$, $A^{A}(v)\ket{\zeta}=\ket{\zeta}$ and $A^{B}(v)\ket{\zeta}=A^{G}(v)\ket{\zeta}=0,$ for all vertices.

\begin{figure}[!t]
    \centering
    \includegraphics[width=0.41\textwidth]{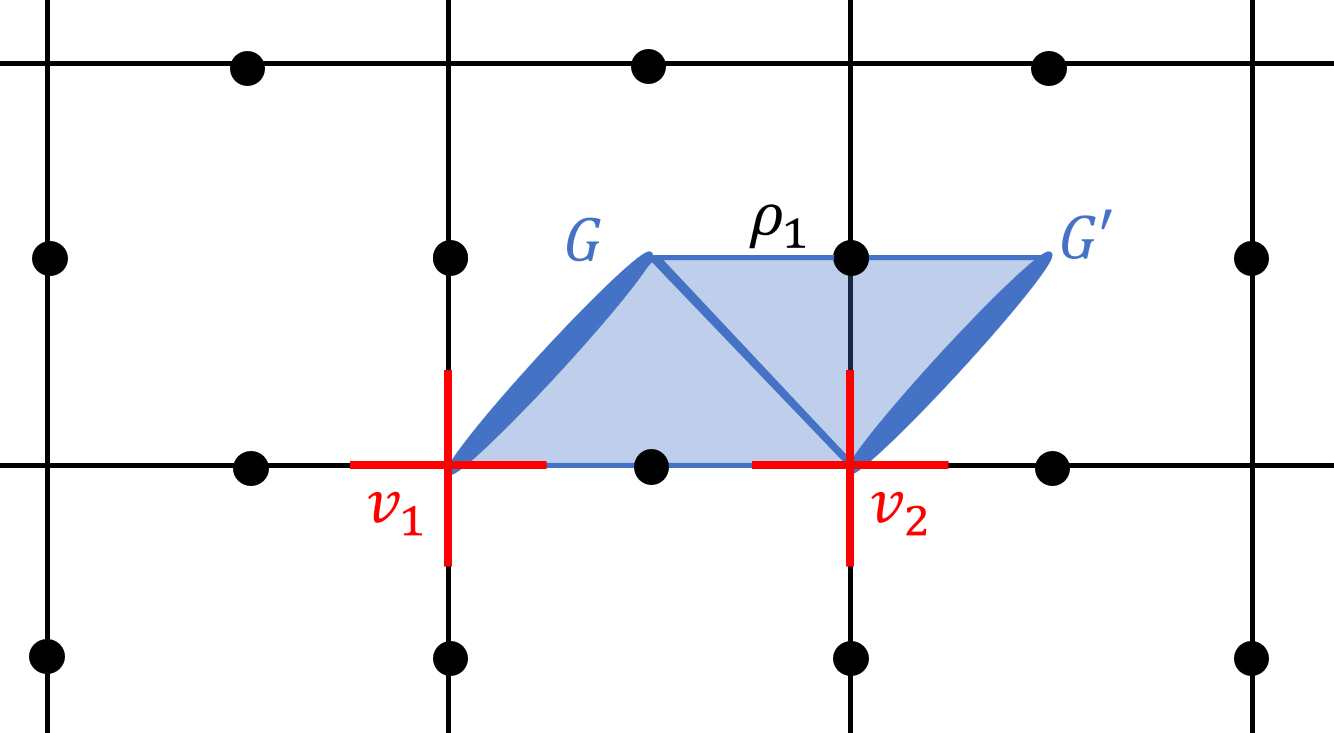}
    \caption{The anyonic ribbon operator $F^G_{\rho_{1}}$ creates a pair of dyons $G$ and $G^{\prime}$ at the endpoints of the ribbon $\rho_{1}$ when acting on the ground state $\ket{\zeta}$. In the $\{A,B,G\}$ sub-model, anyons can be measured with the application of the vertex projection operators such as in \eqref{eq:fusions_of_r1}. In this way, the state $F^{G}_{\rho_{1}}\ket{\zeta}$ is preserved under the action of the measurement operators $A^{G}(v_{1})$ and $A^{G}(v_{2})$.}
    \label{fig:vertex_projection}
\end{figure}

In order to further illustrate the action of these vertex operators consider Figure \ref{fig:vertex_projection}. The ribbon $F^{G}_{\rho_{1}}$ creates a pair of $G$ dyons with sites lying on vertices $v_{1}$ and $v_{2}$. As the charge of the $G$ anyons is unique in the subgroup \{A, B, G\}, it is sufficient to project $G$ anyonic states with the vertex projector $A^G(v)$ as opposed to the full charge and flux projector $P^G$, which requires extra operators on the plaquette to project the flux of the $G$ anyons (see \cite{komar2017anyons}). Application of $A^{G}$ to either vertex will therefore preserve the state, $A^G(v_{1}) F^G_{\rho_1} \ket{\zeta} = F^G_{\rho_1} \ket{\zeta}$, verifying the creation of a $G$ anyon, while $A^{A}(v_{1}) F^G_{\rho_1} \ket{\zeta} = A^{B}(v_{1}) F^G_{\rho_1} \ket{\zeta}  = 0$ (and similarly for $v_{2}$). By employing these projectors on the lattice we can also verify the fusion rules of two $G$ anyons as given in \eqref{eq:fusionrel}. The repeated application of $F^{G}_{\rho_{1}}$ produces a pair of $G$ dyons at both $v_{1}$ and $v_{2}$. Applying a vertex projection operator to either vertex therefore measures the fusion of two $G$ anyons at that point. For example, it is found that
\begin{equation}
    \begin{split}
        & A^A(v_{1}) (F^G_{\rho_1})^2 \ket{\zeta} = \ket{\zeta}, \\
        & A^B(v_{1}) (F^G_{\rho_1})^2 \ket{\zeta} = F^B_{\tau_1} \ket{\zeta}, \\ 
        & A^G(v_{1}) (F^G_{\rho_1})^2 \ket{\zeta} = F^G_{\rho_1} \ket{\zeta}, \label{eq:fusions_of_r1}
    \end{split}
\end{equation}
where $F^B_{\tau_1}$ is the direct triangle ribbon operator producing a pair of $B$ anyons on vertices $v_{1}$ and $v_{2}$, indicating that the fusion of a pair of $G$ anyons gives rise to a superposition of $A$, $B$ and $G$ anyons as dictated by the fusion rule $G\times G =A+B+G$.

\section{R matrix derivation}
\label{sec:R}

In Section \ref{sec:DS3} we introduced the matrices $R^{GG}$ and $F^{G}_{GGG}$, describing the statistical evolutions of the non-Abelian $G$ anyons of the $D(S_{3})$ quantum double model.
The implementation of these evolutions within the framework of the quantum double was first considered by Kitaev in his foundational work \cite{kitaev2003fault}. Here, we begin by presenting such a method by which experimentally accessible ribbon operators may be used to extract the braiding properties of $G$ anyons contained in the matrix elements of $(R^{GG})^{2}$.

\subsection{$R$ matrix from ribbon operators}

\begin{figure}[!t]
    \centering
    \includegraphics[width=0.45\textwidth]{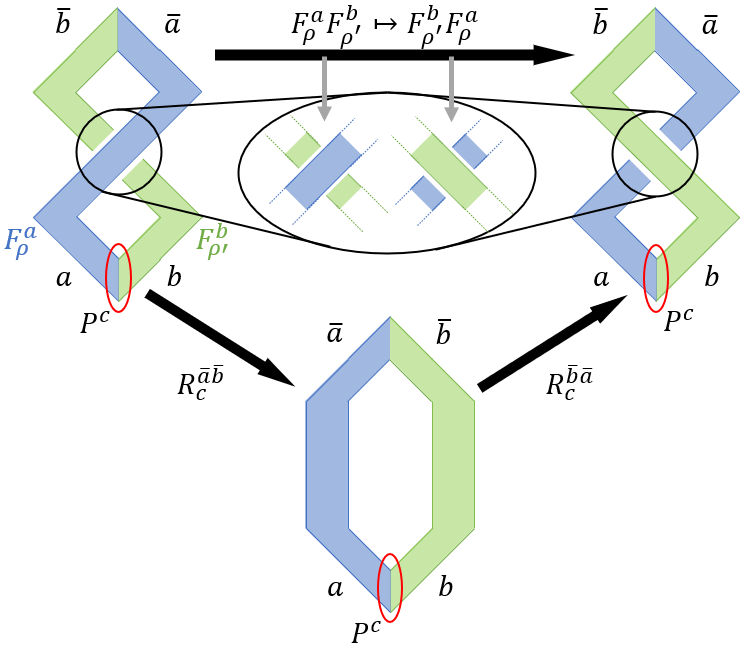}
    \caption{Diagrammatic representation of anyonic braiding by manipulating ribbon operators $F^{a}_{\rho}$ and $F^{b}_{\rho^{\prime}}$ creating pairs of anyons ($a,\bar{a}$) and ($b,\bar{b}$) respectively to obtain the phase $R^{ab}_{c}R^{ba}_{c}$. To highlight the topological equivalence of certain configurations, the ribbons $\rho$ and $\rho^{\prime}$ are abstracted as smoothly deformable strips with no explicit construction on the lattice. 
    In the top line we observe that exchange of order of operation of these two ribbons is equivalent to exchanging the paths on which they overlap.
    The bottom arrows show how to obtain the same configuration by two successive anyonic exchanges, each one characterised by an $R$ phase, giving the relation $P^cF_{\rho}^a F_{\rho^\prime}^b=R^{ab}_{c}R^{ba}_{c}P^cF_{\rho^\prime}^b F^a_{\rho}$.
    }
    \label{fig:rdiag}
\end{figure}

    Consider the ribbon operators $F^{a}_{\rho}$ and $F^{b}_{\rho^{\prime}}$, creating pairs of anyons ($a,\bar{a}$) and ($b,\bar{b}$) respectively as shown in Figure \ref{fig:rdiag}. By considering two equivalent sets of operations connecting the products $F^{a}_{\rho}F^{b}_{\rho^{\prime}}$ and $F^{b}_{\rho^{\prime}}F^{a}_{\rho}$ 
    we will show how the elements of the braiding matrix $R^{ab}$ may be extracted from the commutativity of reduced ribbon operators.
    First, consider exchanging the order of $F^{a}_{\rho}$ and $F^{b}_{\rho^{\prime}}$, as shown in the top line of Figure \ref{fig:rdiag}.
    For all points where the two ribbons do not overlap (indicating action on different qudits on the lattice), such an exchange is trivial. 
    Alternatively, the exchange of these ribbon operators may be interpreted in terms of the projection of the braiding of anyonic worldlines. 
    Both ribbons share common start and end points such that the fusion outcome $a\times b\rightarrow c$ may be fixed with the application of a localised projection operator $P^{c}$ to the point of overlap of $a$ and $b$ (superselection rules then dictate that $\bar{a}\times\bar{b}\rightarrow \bar{c}$ such that the total system will still fuse to the vacuum).
    By analogy with Figure \ref{fig:R_and_F}(a), `undoing' the braiding of the ribbons on the LHS with a clockwise rotation therefore produces any topologically equivalent configuration to that in the centre of Figure \ref{fig:rdiag} with an additional factor $R^{ab}_{c}$. A second clockwise braiding operation brings about another factor of $R^{ab}_{c}$ giving eventually the final configuration on the right. In the anyonic basis such a set of transformations may be represented as
    \begin{equation}
        R^{2}_{\circlearrowleft} \ket{a,b\rightarrow c} = R^{ab}_{c}R^{ba}_{c}  \ket{a,b\rightarrow c}, \label{eq:rabc}
    \end{equation}
    where $R_{\circlearrowleft} \ket{a,b\rightarrow c}$ describes the anti-clockwise exchange of anyons $a$ and $b$ with fixed fusion channel $c$.
    
    By comparison of these two equivalent processes, we therefore observe that the values $R^{ab}_{c}R^{ba}_{c}$ are encoded in the exchange of overlapping ribbon operators as
    \begin{equation}
        P^cF_{\rho}^a F_{\rho^{\prime}}^b=R^{ab}_{c}R^{ba}_{c}P^cF_{\rho^{\prime}}^b F^a_{\rho}.
        \label{eq:Rbraid}
    \end{equation}
    In the following we show how to explicitly reconstruct the matrix $(R^{GG})^{2}$ with operations on the qudit lattice that encodes the $\mathbf{D}(\mathbf{S}_{3})$ quantum double model.

\subsection{Deriving the $R$ matrix on the $\mathbf{D}(\mathbf{S}_{3})$ lattice}
\label{sec:Rprod}

\begin{figure}[!t]
    \centering
    \includegraphics[width=0.45\textwidth]{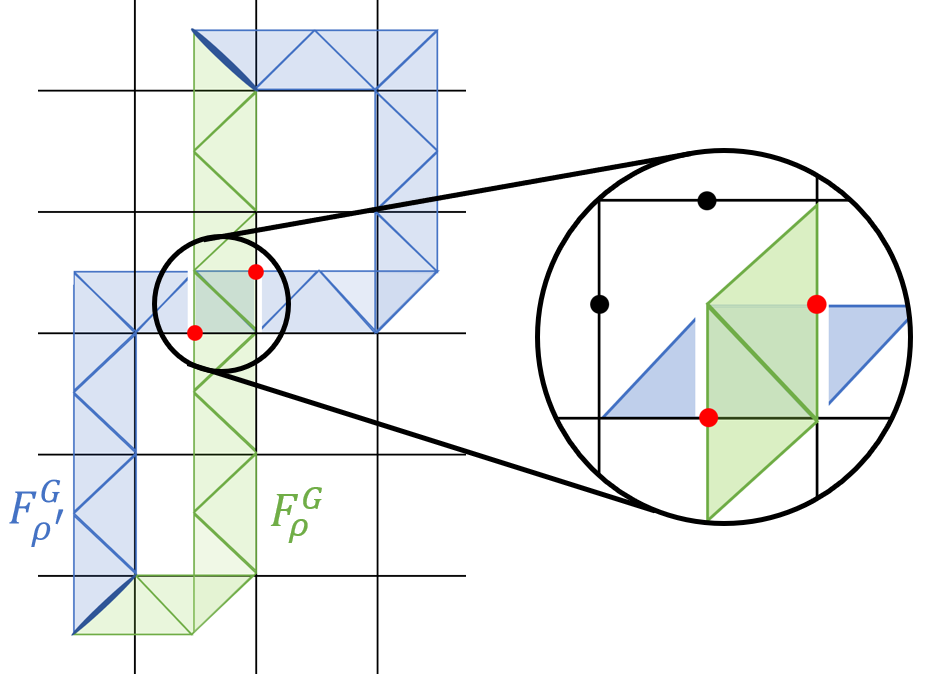}
    \caption{Diagram illustrating the process to obtain the $R^{GG}$ matrix on the quantum double lattice. Following Fig.~\ref{fig:rdiag} we illustrate a pair of braided $G$ anyon ribbons, $F^G_{\rho}$ and $F^G_{\rho'}$, with the same start and end points, highlighting the qudits where the ribbons cross. As the two ribbons, $F^G_{\rho}$ and $F^G_{\rho'}$, cross at two qudits, denoted with red dots, the effect of the braiding only depends on the action on these qudits. Hence, the elements $(R^{GG}_{i})^{2}$ may be reproduced by the reduced system of two simple ribbons, $F^G_{\rho_1}$ and $F^G_{\rho_2}$, as shown in Fig.~\ref{fig:Gs}.}
    \label{fig:Rbraid}
\end{figure}

\begin{figure}[!t]
    \centering
    \includegraphics[width=0.35\textwidth]{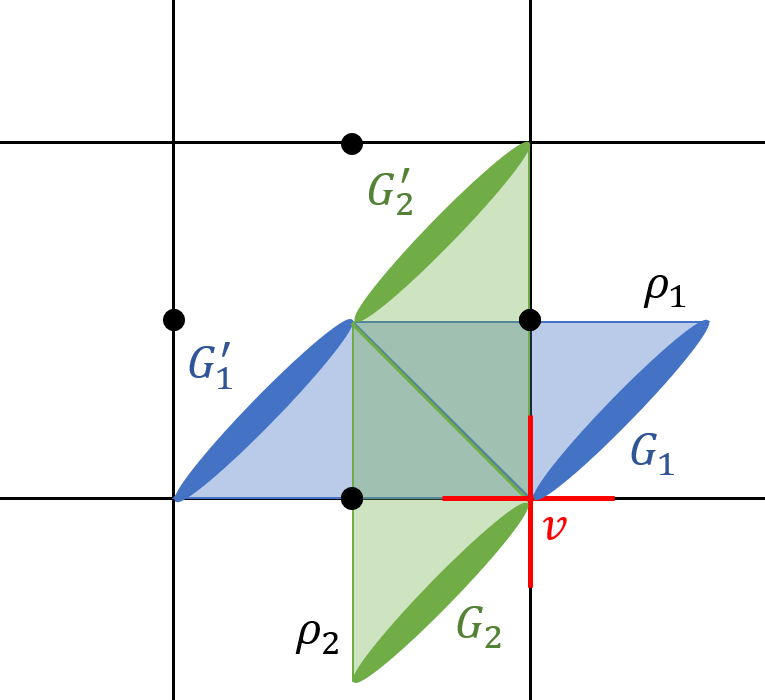}
    \caption{The positions of the four $G$ dyons are depicted as created by the operators $F^{G}_{\rho_{1}}F^{G}_{\rho_{2}}$ and $F^{G}_{\rho_{2}}F^{G}_{\rho_{1}}$. Anyons $G_{1}^{\prime}$ and $G_{2}^{\prime}$ overlap on a plaquette, while $G_{1}$ and $G_{2}$ overlap on the vertex labelled $v$. By applying a vertex projector $A^{i}(v)$, $i=A,B$ or $G$, the outcome of the fusion $G_{1}\times G_{2}$ can be directly measured.}
    \label{fig:Gs}
\end{figure}

Having demonstrated how the braiding of a pair of anyons may be implemented with ribbon operators we now want to show how the abstract diagrams of ribbons in Figure \ref{fig:rdiag} may be explicitly realised on a lattice. 
Consider the ribbons $F^{G}_{\rho}$ and $F^{G}_{\rho'}$ shown in Figure \ref{fig:Rbraid}. In the previous section we have shown how 
topological invariance allows these ribbons to take any arbitrary homotopically equivalent shape as long as the structure of this overlap, which encodes their braiding information, is preserved. 
Hence, the commutativity of two arbitrary ribbons reduces down to the commutativity of the operators acting on these two points of overlap. These minimal operators are $F^{G}_{\rho_{1}}$ and $F^{G}_{\rho_{2}}$ described by \eqref{eq:rho_1} and \eqref{eq:rho_2} respectively and forming a crossed joint on the two qudits as shown in Figure \ref{fig:Gs}. For this system, Equation \eqref{eq:Rbraid} thus becomes
\begin{equation}
    A^{i}(v)F^{G}_{\rho_2} F^{G}_{\rho_1}=(R_i^{GG})^2 A^i(v)F^{G}_{\rho_1} F^G_{\rho_2}
    \label{eq:Gbraid}
\end{equation}
where $A^{i}(v), i=A,B,G,$ is the vertex projector as defined in \eqref{eq:Avproj}, projecting onto the desired fusion outcome of anyons $G_{1}$ and $G_{2}$ as in Figure \ref{fig:Gs}. 

In order to extract the braiding information from these operators we thus introduce two sets of states 
\begin{equation}
    \begin{split}
        \ket{\phi_{21}(i)} = N_{i} A^{i}(v)F^{G}_{\rho_{2}}F^{G}_{\rho_{1}}\ket{\zeta}, \\
        \ket{\phi_{12}(j)} = N_{j} A^{j}(v)F^{G}_{\rho_{1}}F^{G}_{\rho_{2}}\ket{\zeta},
    \end{split}
    \label{eq:1221}
\end{equation}
for $i,j=A,B,G$, with normalisation factors $N_{A}=N_B=2$ and $N_G=\sqrt{2}$ required due to the action of the projectors such that $\langle \phi_{21}(i)|\phi_{21}(i^{\prime})\rangle= \langle \phi_{12}(i)|\phi_{12}(i^{\prime})\rangle =\delta_{ii^{\prime}}$. By analogy with Equation \eqref{eq:Gbraid}, we thus have
\begin{equation}
    \ket{\phi_{21}(i)} = (R^{GG}_{i})^{2}\ket{\phi_{12}(i)}
\end{equation}


such that the algebraic form of each element $(R^{GG})^{i}_{i}\equiv R^{GG}_{i}$ corresponds to the result of the calculation of the overlap
\begin{equation}
    (R^{GG}_{i})^{2} = \langle\phi_{12}(i)|\phi_{21}(i)\rangle.
    \label{eq:Rolap2}
\end{equation}

By explicit computation of these amplitudes we obtain
\begin{equation}
    (R^{GG}_{A})^{2} = (R^{GG}_{B})^{2} = \bar{\omega}, \hspace{0.5cm} (R^{GG}_{G})^{2} = \omega,  
\end{equation}
demonstrating the exact reconstruction of $(R^{GG})^{2}$ in the anyonic basis. Hence, by considering only the two segments $\rho_1$ and $\rho_2$ of the braiding ribbons $\rho$ and $\rho'$ we are able to determine the braiding matrix of the $G$ anyons squared. In Appendix \ref{sec:R_full} we provide an alternative derivation of this result with the evaluation of the operator products $F^{G}_{\rho_{1}}F^{G}_{\rho_{2}}$ and $F^{G}_{\rho_{2}}F^{G}_{\rho_{1}}$ in the basis of group elements.  With comparison of the elements of the resulting matrices, $(R^{GG})^{2}$ may be derived in the anyonic basis in a method lending itself to the quantum tomographic verification of \cite{goel2023unveiling}.

\section{$F$ matrix derivation}

We will now determine the fusion recombination matrix, $F$, illustrated in Figure \ref{fig:R_and_F}(b). Together with the $R$ matrix determined in the previous section, the evaluation of the $F$ matrix will enable us to produce an operator of the $\mathbf{D}(\mathbf{S}_{3})$ braid group and demonstrate their magic state generation capabilities. 
Here we provide a minimal set of ribbon operators and local projections of the planar code model that can determine each of the matrix elements $(F_{GGG}^{G})_{j}^{i}$. 

Surprisingly, the evaluated quantum amplitudes provide the $F$ matrix elements {\em squared}. This `doubling' of fusion processes is obtained when considering the action of ribbon operators on a closed system of anyons, where total charge must be conserved. However, we will show that the obtained information is sufficient to explicitly produce non-Clifford matrices.

\begin{figure*}[!t]
    \centering   
    \includegraphics[width=0.35\textwidth]{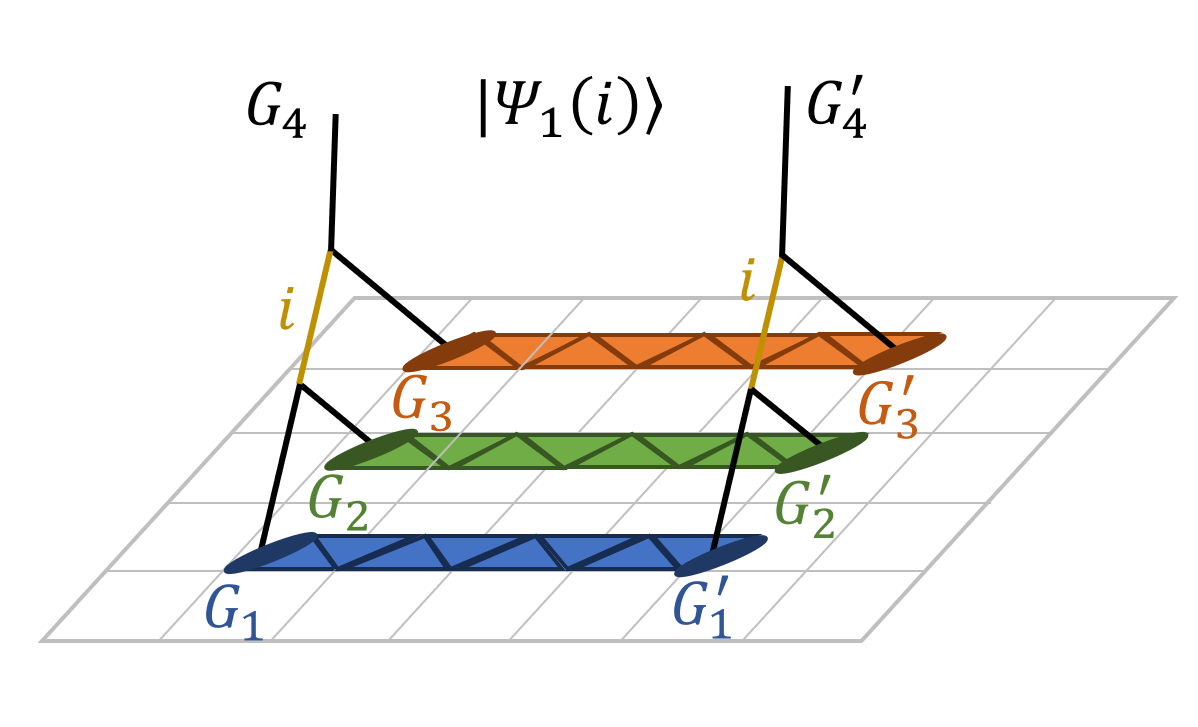}
    \raisebox{0.8cm}{\includegraphics[width=0.175\textwidth]{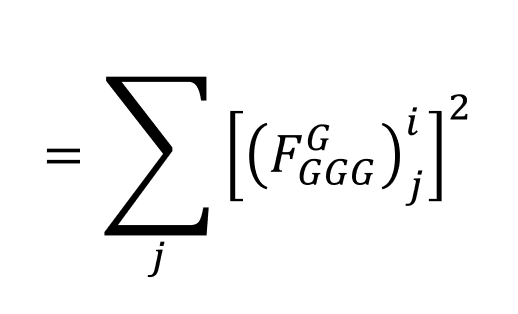}}
    \includegraphics[width=0.35\textwidth]{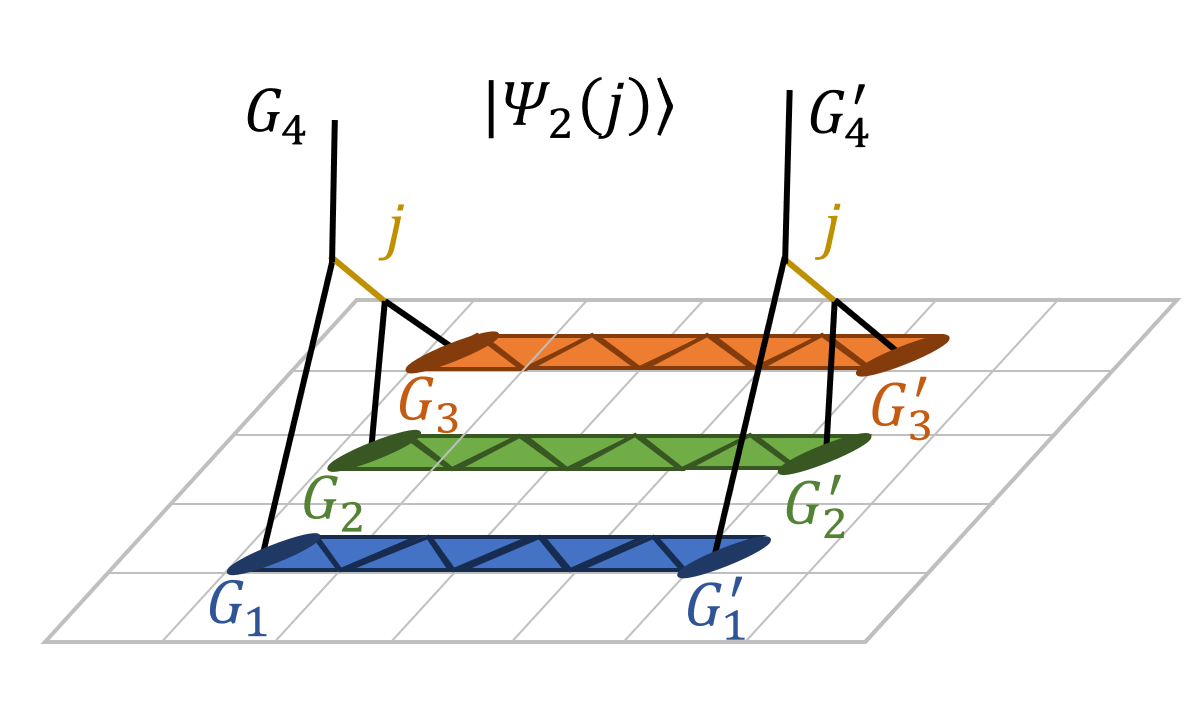}
    \caption{Recombination in the fusion order of $G$ anyons, as seen in Fig. \ref{fig:R_and_F}(b), implemented on the $\mathbf{D}(\mathbf{S}_{3})$ lattice.
    We consider three ribbons $F^{G}_{k}$, $k=1,2,3$ that produce pairs of anyons $G_k$ and $G'_k$ at their endpoints. At one side of the ribbons we measure pairs of anyons in order to impose a certain fusion outcome. Each such measurement necessarily projects the anyons at the other endpoint to the same fusion outcome due to superselection rules. As a result, even if we manipulate only the anyons at one end of the ribbons, the process is doubled and the total state of the system produces squared amplitudes $[(F^G_{GGG})^i_j]^2$ during fusion recombination, as seen in Equation \eqref{eq:FF}.}
    \label{fig:F_double}
\end{figure*}

\subsection{The doubling of fusion recombination}

The matrix $F^{G}_{GGG}$ describes the change of basis in the Hilbert space of four $G$ anyons happening when we exchange the order of fusion of three distinct $G$ anyons, $G_{1},G_{2}, G_{3}$ to a fourth composite $G_4$ anyon. In the notation of Equation \eqref{eq:F} we have
\begin{equation}
    \begin{split}
    \ket{(G_{1},G_{2}), G_3\rightarrow i,G_{3}\rightarrow G_4} \hspace{3cm}\\
    \qquad = 
    \sum_{j} (F^{G}_{GGG})^{i}_{j} \ket{G_1,(G_{2},G_{3})\rightarrow G_{1},j\rightarrow G_4}.
    \end{split}
\end{equation}
Equivalently, the orthogonality of distinct fusion channels means that each matrix element $(F^{G}_{GGG})^{i}_{j}$ may be expressed as the overlap
\begin{equation}
    \begin{split}
    (F^{G}_{GGG})^i_j= \hspace{7cm} \label{eq:Fdef2}  \\
     \langle G_{1},(G_{2},G_{3})\rightarrow G_{1},j \rightarrow G_4|(G_{1},G_{2}),G_{3}\rightarrow i,G_{3}\rightarrow G_4\rangle \nonumber.
    \end{split}
\end{equation}
To successfully simulate this process on the lattice, we therefore want to develop a methodology that allows for the controlled re-ordering of the fusion of three $G$ anyons and the intermediary composite $i$ or $j$. Akin to the methodology for the $R$ matrix, the fusion of $G$ anyons is enacted with the application of combinations of ribbons producing pairs of these anyons $(G_{n},G_{n}^{\prime}), n=1,2,3,4$ as shown in Figure \ref{fig:F_double}. The ordering of fusion is controlled by the application of projective measurements to a chosen pair of anyons. For example, in Fig.~\ref{fig:F_double}(a) the first measurement is of $G_{1}$ and $G_{2}$ to fusion outcome $i$, whereas in Fig.~\ref{fig:F_double}(b), the creation of $j$ is ensured by the projection onto the fusion of $G_{2}$ and $G_{3}$. We note that each of our states is constructed from the application of ribbon operators to the vacuum. Superselection therefore dictates that at each stage it must always be possible to fuse the created anyons back to the vacuum. As each of the ribbon operators in the quantum double necessarily create anyons in pairs, we will see that anyonic conservation therefore results in two identical fusion processes occurring simultaneously between two sets of anyons, $(G_1,G_2, G_3, G_4)$ and $(G'_1,G'_2, G'_3, G'_4)$ as shown in Figure \ref{fig:F_double}. 

To make this analysis explicit, we denote the two sets of identical fusion processes with the tensor product such that in the anyonic basis we obtain
\begin{equation}
    \begin{split}
        \ket{\Psi_{1}(i)} = \ket{(G_{1},G_{2}),G_{3}\rightarrow i,G_{3}\rightarrow G} \\
        \otimes \ket{(G_{1}^{\prime},G_{2}^{\prime}),G_{3}^{\prime}\rightarrow i,G_{3}^{\prime}\rightarrow G}
    \end{split}
\end{equation}
and
\begin{equation}
    \begin{split}
        \ket{\Psi_{2}(j)} = \ket{G_{1},(G_{2},G_{3})\rightarrow G_{1},j\rightarrow G} \\
        \otimes \ket{G_{1}^{\prime},(G_{2}^{\prime},G_{3}^{\prime})\rightarrow G_{1}^{\prime},j\rightarrow G}.
    \end{split}
\end{equation}
It is important here to note that although the matrix $F^{G}_{GGG}$ relates two sets of orthonormal basis states as in Equation \eqref{eq:F}, this doubling of fusion processes produced by projection operators is described by the non-unitary matrix with elements $\mathcal{F}^{i}_{j} = [\left(F^{G}_{GGG}\right)^{i}_{j}]^{2}$. In order to extract these elements it is therefore necessary to choose one set of states as an orthonormal basis. The set of states created by re-ordering the fusion processes as detailed may then not be orthonormal, but can be expressed as a linear combination of the other set of states. For example, if $\ket{\Psi_{2}(j)}$
is constructed as the orthonormal basis, then 
\begin{equation}
    \ket{\Psi_{1}(i)} = \sum_{j} [\left(F^{G}_{GGG}\right)^{i}_{j}]^{2} \ket{\Psi_{2}(j)}
    \label{eq:FF}
\end{equation}
In this way we obtain
\begin{equation}
    \begin{split}
        \langle \Psi_{2}(j)|\Psi_{1}(i)\rangle = \left[(F^{G}_{GGG})^{i}_{j}\right]^{2},
    \end{split}
    \label{eq:psi}
\end{equation}
where one $(F^{G}_{GGG})^{i}_{j}$ comes from the recombination of the unprimed $G$ anyons and the other from the primed ones. Such a method for reproducing the $F$ matrix using a closed system will therefore necessarily produce the squared value of each element. This is a general result differing from the case of the braiding procedure we saw in Section \ref{sec:Rprod} that produces $(R^{GG})^2$, which is due to the minimal size of the ribbons we consider.
Below, we will introduce a set of states that explicitly demonstrate these anyonic processes with the quantum double lattice model. 

\subsection{Deriving the $F$ matrix on the $\mathbf{D}(\mathbf{S}_{3})$ lattice}

Here we introduce a method to explicitly determine the elements
\begin{equation}
    \mathcal{F}^{i}_{j} = \left[\left(F^{G}_{GGG}\right)^{i}_{j}\right]^{2}.
\end{equation}
of the `squared' matrix $\mathcal{F}$, by considering operations on the lattice of the $\mathbf{D}(\mathbf{S}_{3})$ quantum double model. Our main aim is to determine these matrix elements with the minimal amount of resources. In particular, we use the two qudit ribbons $F^{G}_{\rho_{1}}$ and $F^{G}_{\rho_{2}}$ defined in \eqref{eq:rho_1} and \eqref{eq:rho_2}. As these ribbons overlap extra care must be taken to account for possible braiding phase factors, as we shall see in the following. We start by introducing two states
\begin{equation}
    \begin{split}
        & \ket{\psi_{1}(i)} = A^{G_{4}}(v) F^{G_{3}}_{\rho_{2}} A^{i}(v) F^{G_{2}}_{\rho_{2}} F^{G_{1}}_{\rho_{1}} \ket{\zeta}, \\
        & \ket{\psi_{2}(j)} = A^{G_{4}}(v) F^{G_{1}}_{\rho_{1}} A^{j}(v) F^{G_{3}}_{\rho_{2}} F^{G_{2}}_{\rho_{2}} \ket{\zeta}, \label{eq:psi_states_}
    \end{split}
\end{equation}
for $i,j=A,B,G$. The subscripts in $G_n$ are provided here to elucidate the link to the process shown in Figure \ref{fig:F_double}, but will be dropped in the following when the distinction is not needed. Both states $\ket{\psi_{1}(i)}$ and $\ket{\psi_{2}(j)}$ are not normalised due to the application of the projectors $A^{i}(v)$. Nevertheless, states $\ket{\psi_2(j)}$ are orthogonal so in the following we account explicitly for their normalisation, $\braket{\psi_2(j) | \psi_2(j)}$. The states $\ket{\psi_1(i)}$ are not orthogonal to each other, but similar to \eqref{eq:FF} they can be represented as superpositions of $\ket{\psi_2(j)}$ states as
\begin{equation}
    \ket{\psi_1(i)} = f_{iA} \ket{\psi_2(A)} + f_{iB} \ket{\psi_2(B)} + f_{iG} \ket{\psi_2(G)}. \label{eq:psi_superposition}
\end{equation}
As $\ket{\psi_2(j)}$'s are orthogonal we thus have
\begin{equation}
    f_{ij}  = \frac{\braket{\psi_2(j) | \psi_1(i)}}{\braket{\psi_2(j) | \psi_2(j)}}. \label{eq:fff}
\end{equation}
Similar to \eqref{eq:psi}, the coefficients $f_{ij}$ are related to the elements of the matrix $\mathcal{F}$ up to overall phase factors generated due to the crossing of the $\rho_1$ and $\rho_2$ ribbons.  

In the states $\ket{\psi_1(i)}$ and $\ket{\psi_2(j)}$, the order of fusion is explicitly enforced with the order of application of each of the ribbon operators, $F_{\rho_1}^G$ and $F_{\rho_2}^G$ to the ground state $\ket{\zeta}$. Consider $\ket{\psi_{1}(i)}$, the initial action of $F^{G}_{\rho_{2}}F^{G}_{\rho_{1}}$ on the anyonic vacuum $\ket{\zeta}$ produces two pairs of $G$ anyons as illustrated in Figure \ref{fig:Gs}. By applying the projector $A^{i}(v)$ to the vertex $v$ on which two of these anyons from different ribbons overlap, we may ensure the fusion outcome $G_{1}\times G_{2}\rightarrow i$. Superselection also ensures the simultaneous complementary fusion process $G_{1}^{\prime}\times G_{2}^{\prime}\rightarrow i$. The second application of $F^{G}_{\rho_{2}}$ to this state produces a subsequent pair of fusion processes in which the anyons $G_{3}$ and $G_{3}^{\prime}$ fuse with the composite $i$ anyons. The application of the final projector $A^{G}(v)$ to this vertex ensures that both $G_{3}\times i\rightarrow G_{4}$ and $G_{3}^{\prime}\times i\rightarrow G_{4}^{\prime}$, thus completing the construction of the fusion trees as shown in Figure \ref{fig:F_double}. Note that as we require the use of a small system to realise the fusion properties, the ribbon operators used in $\ket{\psi_{1}(i)}$ and $\ket{\psi_{2}(j)}$ are crossing each other, giving rise to additional braiding phase factors. 

With analytic consideration of the overlaps $\langle\psi_{2}(j)|\psi_{1}(i)\rangle$ as shown in Appendix \ref{sec:Folap}, we find that it is always possible to extract a phase factor $(R^{Gj}_{G})^2$ due to the braiding of the worldlines of the $j$ and $G_{3}$ anyons in $\ket{\psi_{2}(j)}$. In particular, we have
\begin{equation}
    \begin{split}
        & \braket{\psi_2(j) | \psi_1(i)} = (R^{Gj}_G)^2 \bra{\zeta} F^{G}_{\rho_{1}} F^{j}_{\rho_{2}} A^{G}(v) F^{G}_{\rho_{2}} A^{i}(v) F^{G}_{\rho_{2}} F^{G}_{\rho_{1}} \ket{\zeta}.
    \end{split}
    \label{eq:FRphase}
\end{equation}

We therefore find that, by using \eqref{eq:psi_states_} and \eqref{eq:psi_superposition}, the fusion matrix elements squared may be calculated from
\begin{equation}
    \begin{split}
        \mathcal{F}^i_j & =  {\overline{(R^{Gj}_{G}})^2} f_{ij},
        \label{eq:Ffull}
    \end{split}
\end{equation}
where the phases
\begin{equation}
    (R^{GA}_{G})^2=(R^{GB}_{G})^2=1, \hspace{0.5cm} (R^{GG}_{G})^2=\omega.
\end{equation}
Calculation of these overlaps thus yields the `squared' $F$ matrix
\begin{equation}
    \mathcal{F} = \frac{1}{4}\begin{pmatrix}
        1 & 1 & 2 \\
        1 & 1 & 2 \\
        2 & 2 & 0
    \end{pmatrix}.
\end{equation}
This result is in agreement with the $F$-matrix of the $G$ anyons of $\mathbf{D}(\mathbf{S}_{3})$ given in \eqref{eq:F_matrix}, as determined from the pentagon equations. As this matrix is real and positive the values of the $F$ matrix elements can be derived up to a sign ambiguity, due to the doubling of the fusion processes when ribbons are employed in the fusion recombination, as described in Fig. \ref{fig:F_double}. In the following we show that the $R^{2}$ and $F$ matrices can still produce the non-Clifford unitaries required to produce quantum magic.

\subsection{Non-Clifford action of $B_{2,G}$}

In the previous we employed manipulations on the lattice of the $\mathbf{D}(\mathbf{S}_{3})$ quantum double model to determine the elements of the fusion matrix, $F$. As the method we employed gives rise to squares of the elements, $\mathcal{F}^{i}_{j} = \left[\left(F^{G}_{GGG}\right)^{i}_{j}\right]^{2}$, a sign ambiguity arises in determining the elements of $F^{G}_{GGG}$ when extracted from $\mathcal{F}$. It is possible however, to unequivocally show that for any valid combination of signs, the corresponding braiding matrix always encodes a logical non-Clifford gate capable of producing magic states, which is the goal of this investigation.

To proceed we consider all matrices $\tilde{F}^{G }_{GGG}$ that satisfy
\begin{equation}
   \mathcal{F}^{i}_{j} = \left[\left(\tilde F^{G}_{GGG}\right)^{i}_{j}\right]^{2}.
    \label{eq:relation}
\end{equation}
Accounting for sign ambiguity, this set of matrices may be parametrised as
\begin{equation}
    \left\{\widetilde F^{G}_{GGG}\right\} = \left\{\frac{1}{2} \begin{pmatrix}
        \alpha & \beta & \gamma\sqrt{2} \\
        \delta & \epsilon & \zeta\sqrt{2} \\
        \eta\sqrt{2} & \kappa\sqrt{2} & 0
    \end{pmatrix}, \alpha,\beta,\dots,\kappa = \pm 1\right\},
    \nonumber
\end{equation}
where the additional restraint of unitarity reduces the number of valid solutions to sixty-four matrices. Utilising the extracted form of $(R^{GG})^{2}$ and Eq.~\eqref{eq:B2G}, a braiding matrix $\widetilde{B}_{2,G}=\widetilde F^{G\hspace{0.1cm}-1}_{GGG}(R^{GG})^{2}\widetilde F^{G}_{GGG}$ may be computed for each such solution. Remarkably, it is found that even with this sign ambiguity, all resulting braiding operations may be expressed as
\begin{equation}
    \widetilde{B}_{2,G} = \begin{pmatrix}
        \cos{\left(\frac{2\pi}{3}\right)} & \pm i\sin{\left(\frac{2\pi}{3}\right)} & 0\\
        \pm i\sin{\left(\frac{2\pi}{3}\right)} & \cos{\left(\frac{2\pi}{3}\right)} & 0 \\
        0 & 0 & \bar{\omega}
    \end{pmatrix},
\end{equation}
explicitly verifying the non-Clifford action of this braiding on the reduced fusion subspace $\text{span}(\ket{A},\ket{B})$.
In particular, the action of $\widetilde{B}_{2,G}$ on the intermediate vacuum fusion state $\ket{A}$ yields two possible states 
\begin{equation}
    \ket{\psi_{\pm}} = \cos{\left(\frac{2\pi}{3}\right)}\ket{A} \pm i\sin{\left(\frac{2\pi}{3}\right)}\ket{B}.
\end{equation}
To determine if these states can be used as magic states we evaluate their second order Renyi entropy, $M_{2}$. Using~\eqref{eq:SRE} we find that
each has a non-zero stabilizer Renyi entropy $M_{2}(\ket{\psi_{\pm}})=\log{\left(\frac{16}{13}\right)}$. In this way, the simple lattice protocol presented in this section, provides an explicit demonstration of the realisation of valuable magic states through the braiding of non-Abelian $G$ anyons.

\section{Reconstructing $R$ and $F$ matrices with two $d=6$ qudits}

Our considerations thus far have been applicable to a quantum double lattice of arbitrary size. 
Here we will show that, by appropriate state preparation and  manipulation, the whole process of determining $(R^{GG})^2$ and $F^{G}_{GGG}$ described above can be encoded on two qudits only.

\begin{figure}[!t]
    \centering
    \includegraphics[width=0.4\textwidth]{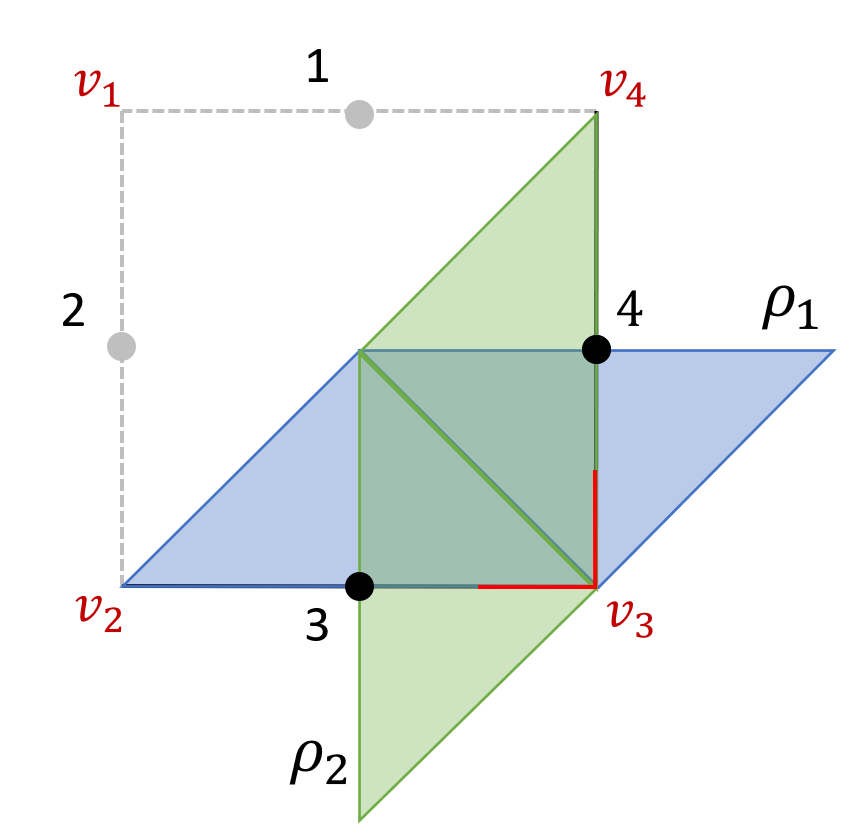}
    \caption{
    The Kitaev Hamiltonian \eqref{eq:H} defined on a single plaquette with open boundary conditions. In this system, both the ribbon operators $F^{G}_{\rho_{1}}$ and $F^{G}_{\rho_{2}}$ and projector $A^{i}(v_{3})$ have support on only qudits $3$ and $4$, facilitating a dense encoding of the lattice reconstruction of the $R$ and $F$  matrices.}
    \label{fig:half_plaquette}
\end{figure}

We start by considering the minimal lattice composed of four qudits on a single plaquette, $p$, with open boundary conditions as shown in Fig.~\ref{fig:half_plaquette}. The Hamiltonian of the system is given by
\begin{equation}
\mathcal{H}_{4} = -B(p) - \sum_{i=1}^{4} A(v_{i}),
\label{eq:H4}
\end{equation}
where the vertex operators, $A(v_{i})$, act non-trivially on only two qudits. These reduced vertex projection operators act as boundary terms in the Hamiltonian \eqref{eq:H4}. We note that the choice of boundary conditions for this minimal case is made without loss of generality but only to ensure that the corresponding model $\mathcal{H}_{4}$ has a unique ground state, $\ket{\eta}$, that has support on all four qudits. When creating pairs of $G$ dyons with the application of $F^{G}_{\rho_{1}}$ and $F^{G}_{\rho_{2}}$ to $\ket{\eta}$, only a single flux is supported by this plaquette, the flux of the $G$ dyon at the other end of the ribbon will lie outside of the system and thus cannot be directly measured.

In Equations \eqref{eq:Rolap2} and \eqref{eq:Ffull} we have shown that the derivation of the forms of both the $R$ and $F$ matrices can be expressed in terms of the overlap between the generalised states $\ket{\phi_{12}(i)}$ \& $\ket{\phi_{21}(j)}$ and $\ket{\psi_{1}(i)}$ \& $\ket{\psi_{2}(j)}$, respectively. For the case of a single plaquette system the states $\ket{\phi_{12}(i)}$ and $\ket{\phi_{21}(j)}$ are given by
\begin{equation}
    \begin{split}
        & \ket{\phi_{12}(i)} = N_i A^{i}(v_{3}) F^{G}_{\rho_{1}} F^{G}_{\rho_{2}} \ket{\eta}, \\
        & \ket{\phi_{21}(j)} = N_j A^{j}(v_{3}) F^{G}_{\rho_{2}} F^{G}_{\rho_{1}} \ket{\eta}, \label{eq:psi_states2}
    \end{split}
\end{equation}
while $\ket{\psi_{1}(i)}$ and $\ket{\psi_{2}(j)}$ become
\begin{equation}
    \begin{split}
        & \ket{\psi_{1}(i)} = A^{G}(v_{3}) F^{G}_{\rho_{2}} A^{i}(v_{3}) F^{G}_{\rho_{2}} F^{G}_{\rho_{1}} \ket{\eta}, \\
        & \ket{\psi_{2}(j)} = A^{G}(v_{3}) F^{G}_{\rho_{1}} A^{j}(v_{3}) F^{G}_{\rho_{2}} F^{G}_{\rho_{2}} \ket{\eta}. \label{eq:psi_states3}
    \end{split}
\end{equation}
The ribbon operators $F^{G}_{\rho_{1}}$ and $F^{G}_{\rho_{2}}$ act as defined in Equations \eqref{eq:rho_1} and \eqref{eq:rho_2} on qudits $3$ and $4$, as shown in Figure~\ref{fig:half_plaquette}. Similarly, the projector $A^G(v_{3})$ also acts on qudits 3 and 4, as seen in \eqref{eqn:pro}. Hence, every operator in the amplitudes $\langle\phi_{21}(i)|\phi_{12}(i)\rangle$ and $\braket{\psi_2(j) | \psi_1(i)}$ only acts non-trivially on two qudits. It can be shown (see Appendix \ref{sec:2qudits}) that for any operator, $O$, that acts only on qudits 3 and 4, its expectation value with respect to the four qudit ground state $\ket{\eta}$, can be given in terms of the summation of six independent expectation values
\begin{equation}
\langle \eta |\mathbf{1}_{36}\otimes O |\eta \rangle = 6\sum_{g\in G} \bra{\psi_{g}}O\ket{\psi_{g}},
\label{eqn:simple}
\end{equation}
where
\begin{equation}
    \ket{\psi_{g}} = \frac{1}{\sqrt{216}}\sum_{g_{1}g_{2}=g} \ket{g_{1},g_{2}}.
    \label{eq:xi}
\end{equation}
Here, each $\ket{\psi_{g}}$ is a 36-dimensional column vector that, similarly to $O$, has support on only the two qudits $3$ and $4$. Employing \eqref{eqn:simple} and \eqref{eq:xi} one can directly determine $(R^{GG}_{i})^2 = \langle\phi_{21}(i)|\phi_{12}(i)\rangle$
and $[(F^{G}_{GGG})^{i}_{j}]^{2}=(\overline{R^{Gj}_{G}})^2\langle\psi_{2}(j)|\psi_{1}(i)\rangle$ by using only two $d=6$ qudits, obtaining the same results as the full system. Hence, although the minimal physical system for implementing such an anyonic model is a single plaquette on four qudits, the ground state structure and reduced action of associated operators allows for the same information on the $R$ and $F$ matrix structure to be extracted from the overlaps of states constructed on two qudits.

\section{Conclusion}

This work represents a significant step towards the realisation of non-Abelian anyons in practical quantum systems, offering a clear pathway for experimental demonstration and potential applications in universal quantum computation and error correction. For this task, we considered the $\mathbf{D}(\mathbf{S}_{3})$ quantum double model, which supports the closed anyonic subgroup $\{A,B,G\}$, with $G$ being the non-Abelian anyon. This model can be encoded on a square lattice of $d=6$ qudits. While generating and manipulating $G$ anyons in this model typically requires a lattice of several qudits and controlled operations between them, we have shown that by carefully investigating the braiding and fusion properties of the $G$ anyons, this information can be decoded from the ribbons that generate them using simple projective measurements. Furthermore, we developed a scheme whereby the $R$ and $F$ matrices of the $G$ anyons can be determined using only two qudits. This significantly reduces the experimental resources required to demonstrate their ability to generate magic states capable for universal quantum computation. Moreover, by compactifying the fundamental building block to just two qudits, we enable the direct scalability of the system to multiple anyons, where larger encoded Hilbert spaces can be realised, paving the way for the execution of prototype quantum algorithms. 

Physical implementations of non-Abelian anyonic systems present significant experimental challenges due to the need for precise control of complex quantum systems. Our results establish that the fundamental operations, such as state preparation, braiding, fusion, and measurement, can be implemented with minimal resources, charting a route toward universal quantum computation based on non-Abelian $\mathbf{D}(\mathbf{S}_{3})$ anyons. To that end, various platforms, including superconducting circuits, trapped ions, and cold atoms, can be employed to simulate the desired anyonic evolutions, each with distinct advantages and limitations. In particular, photonic systems have emerged as a powerful candidate for simulating non-Abelian anyons due to their inherent ability to implement high-fidelity projective measurements and non-unitary operations. As demonstrated in previous work \cite{goel2023unveiling}, photonic platforms allow the encoding of anyonic degrees of freedom in high-dimensional spatial modes, where non-Abelian fusion and braiding matrices can be implemented with high precision using spatial light modulators and mode-mixing elements. We foresee that the experimental implementation of our minimal system will significantly contribute to the development and realisation of non-Abelian quantum error-correcting codes, which offer high error thresholds and minimal resource requirements, thereby advancing the field of fault-tolerant quantum computation.\\

\acknowledgements

This work was supported by EPSRC with Grants
EP/R020612/1 and UKRI1337:Anyons24. L.B. acknowledges support from EPSRC Grant No. EP/T517860/1.

\bibliographystyle{ieeetr}
\bibliography{refs}

@article{mochon2004anyon,
  title={Anyon computers with smaller groups},
  author={Mochon, Carlos},
  journal={Physical Review A},
  volume={69},
  number={3},
  pages={032306},
  year={2004},
  publisher={APS}
}

@article{wootton2014error,
  title={Error correction for non-abelian topological quantum computation},
  author={Wootton, James R and Burri, Jan and Iblisdir, Sofyan and Loss, Daniel},
  journal={Physical review X},
  volume={4},
  number={1},
  pages={011051},
  year={2014},
  publisher={APS}
}

@article{luo2011simulation,
  title={Simulation of non-Abelian anyons using ribbon operators connected to a common base site},
  author={Luo, Xi-Wang and Han, Yong-Jian and Guo, Guang-Can and Zhou, Xingxiang and Zhou, Zheng-Wei},
  journal={Physical Review A},
  volume={84},
  number={5},
  pages={052314},
  year={2011},
  publisher={APS}
}

@article{komar2017anyons,
  title={Anyons are not energy eigenspaces of quantum double Hamiltonians},
  author={K{\'o}m{\'a}r, Anna and Landon-Cardinal, Olivier},
  journal={Physical Review B},
  volume={96},
  number={19},
  pages={195150},
  year={2017},
  publisher={APS}
}

@article{goel2023unveiling,
  title={Unveiling the non-Abelian statistics of $ D (S\_3) $ anyons via photonic simulation},
  author={Goel, Suraj and Reynolds, Matthew and Girling, Matthew and McCutcheon, Will and Leedumrongwatthanakun, Saroch and Srivastav, Vatshal and Jennings, David and Malik, Mehul and Pachos, Jiannis K},
  journal={arXiv preprint arXiv:2304.05286},
  year={2023}
}

@article{bombin2008family,
  title={Family of non-Abelian Kitaev models on a lattice: Topological condensation and confinement},
  author={Bombin, Hector and Martin-Delgado, MA},
  journal={Physical Review B},
  volume={78},
  number={11},
  pages={115421},
  year={2008},
  publisher={APS}
}

@article{kitaev2003fault,
  title={Fault-tolerant quantum computation by anyons},
  author={Kitaev, A Yu},
  journal={Annals of physics},
  volume={303},
  number={1},
  pages={2--30},
  year={2003},
  publisher={Elsevier}
}

@book{pachos2012introduction,
  title={Introduction to topological quantum computation},
  author={Pachos, Jiannis K},
  year={2012},
  publisher={Cambridge University Press}
}

@article{laubscher2019universal,
  title={Universal quantum computation in the surface code using non-Abelian islands},
  author={Laubscher, Katharina and Loss, Daniel and Wootton, James R},
  journal={Physical Review A},
  volume={100},
  number={1},
  pages={012338},
  year={2019},
  publisher={APS}
}

@book{simon2023topological,
  title={Topological quantum},
  author={Simon, Steven H},
  year={2023},
  publisher={Oxford University Press}
}

@article{xu2016simulating,
  title={Simulating the exchange of Majorana zero modes with a photonic system},
  author={Xu, Jin-Shi and Sun, Kai and Han, Yong-Jian and Li, Chuan-Feng and Pachos, Jiannis K and Guo, Guang-Can},
  journal={Nature communications},
  volume={7},
  number={1},
  pages={13194},
  year={2016},
  publisher={Nature Publishing Group UK London}
}

@article{liu2021topological,
  title={Topological contextuality and anyonic statistics of photonic-encoded parafermions},
  author={Liu, Zheng-Hao and Sun, Kai and Pachos, Jiannis K and Yang, Mu and Meng, Yu and Liao, Yu-Wei and Li, Qiang and Wang, Jun-Feng and Luo, Ze-Yu and He, Yi-Fei and others},
  journal={PRX Quantum},
  volume={2},
  number={3},
  pages={030323},
  year={2021},
  publisher={APS}
}

@article{aguado2008creation,
  title={Creation, manipulation, and detection of Abelian and non-Abelian anyons in optical lattices},
  author={Aguado, Miguel and Brennen, GK and Verstraete, Frank and Cirac, J Ignacio},
  journal={Physical review letters},
  volume={101},
  number={26},
  pages={260501},
  year={2008},
  publisher={APS}
}

@article{moore1991nonabelions,
  title={Nonabelions in the fractional quantum Hall effect},
  author={Moore, Gregory and Read, Nicholas},
  journal={Nuclear Physics B},
  volume={360},
  number={2-3},
  pages={362--396},
  year={1991},
  publisher={Elsevier}
}

@article{dolev2008observation,
  title={Observation of a quarter of an electron charge at the $\nu$= 5/2 quantum Hall state},
  author={Dolev, Merav and Heiblum, Moty and Umansky, V and Stern, Ady and Mahalu, Diana},
  journal={Nature},
  volume={452},
  number={7189},
  pages={829--834},
  year={2008},
  publisher={Nature Publishing Group UK London}
}

@article{bacon2008stability,
  title={Stability of quantum concatenated-code Hamiltonians},
  author={Bacon, Dave},
  journal={Physical Review A—Atomic, Molecular, and Optical Physics},
  volume={78},
  number={4},
  pages={042324},
  year={2008},
  publisher={APS}
}

@article{li2024photonic,
  title={Photonic simulation of Majorana-based Jones polynomials},
  author={Li, Jia-Kun and Sun, Kai and Hao, Ze-Yan and Liang, Jia-He and Tao, Si-Jing and Pachos, Jiannis K and Xu, Jin-Shi and Han, Yong-Jian and Li, Chuan-Feng and Guo, Guang-Can},
  journal={arXiv preprint arXiv:2403.04980},
  year={2024}
}

@article{weimer2010rydberg,
  title={A Rydberg quantum simulator},
  author={Weimer, Hendrik and M{\"u}ller, Markus and Lesanovsky, Igor and Zoller, Peter and B{\"u}chler, Hans Peter},
  journal={Nature Physics},
  volume={6},
  number={5},
  pages={382--388},
  year={2010},
  publisher={Nature Publishing Group UK London}
}

@article{benhemou2023universality,
  title={Universality of Z 3 parafermions via edge-mode interaction and quantum simulation of topological space evolution with Rydberg atoms},
  author={Benhemou, Asmae and Angkhanawin, Toonyawat and Adams, Charles S and Browne, Dan E and Pachos, Jiannis K},
  journal={Physical Review Research},
  volume={5},
  number={2},
  pages={023076},
  year={2023},
  publisher={APS}
}

@article{raussendorf2007topological,
  title={Topological fault-tolerance in cluster state quantum computation},
  author={Raussendorf, Robert and Harrington, Jim and Goyal, Kovid},
  journal={New Journal of Physics},
  volume={9},
  number={6},
  pages={199},
  year={2007},
  publisher={IOP Publishing}
}

@article{nielsen2004optical,
  title={Optical quantum computation using cluster states},
  author={Nielsen, Michael A},
  journal={Physical review letters},
  volume={93},
  number={4},
  pages={040503},
  year={2004},
  publisher={APS}
}

@article{nayak2008non,
  title={Non-Abelian anyons and topological quantum computation},
  author={Nayak, Chetan and Simon, Steven H and Stern, Ady and Freedman, Michael and Das Sarma, Sankar},
  journal={Reviews of Modern Physics},
  volume={80},
  number={3},
  pages={1083--1159},
  year={2008},
  publisher={APS}
}

@article{das2005topologically,
  title={Topologically protected qubits from a possible non-Abelian fractional quantum Hall state},
  author={Das Sarma, Sankar and Freedman, Michael and Nayak, Chetan},
  journal={Physical review letters},
  volume={94},
  number={16},
  pages={166802},
  year={2005},
  publisher={APS}
}

@article{read2000paired,
  title={Paired states of fermions in two dimensions with breaking of parity and time-reversal symmetries and the fractional quantum Hall effect},
  author={Read, Nicholas and Green, Dmitry},
  journal={Physical Review B},
  volume={61},
  number={15},
  pages={10267},
  year={2000},
  publisher={APS}
}

@article{ivanov2001non,
  title={Non-Abelian statistics of half-quantum vortices in p-wave superconductors},
  author={Ivanov, Dmitri A},
  journal={Physical review letters},
  volume={86},
  number={2},
  pages={268},
  year={2001},
  publisher={APS}
}

@article{yu2020non,
  title={Non-Majorana states yield nearly quantized conductance in superconductor-semiconductor nanowire devices},
  author={Yu, P and Chen, J and Gomanko, M and Badawy, G and Bakkers, EPAM and Zuo, K and Mourik, V and Frolov, SM},
  journal={arXiv preprint arXiv:2004.08583},
  year={2020}
}

@article{google2023non,
  author={Andersen, Trond I, et al.},
  title={Non-Abelian braiding of graph vertices in a superconducting processor},
  journal={Nature},
  volume={618},
  number={7964},
  pages={264--269},
  year={2023},
  publisher={Nature Publishing Group UK London}
}

@article{stenger2021simulating,
  title={Simulating the dynamics of braiding of Majorana zero modes using an IBM quantum computer},
  author={Stenger, John PT and Bronn, Nicholas T and Egger, Daniel J and Pekker, David},
  journal={Physical Review Research},
  volume={3},
  number={3},
  pages={033171},
  year={2021},
  publisher={APS}
}

@article{wootton2017demonstrating,
  title={Demonstrating non-Abelian braiding of surface code defects in a five qubit experiment},
  author={Wootton, James R},
  journal={Quantum Science and Technology},
  volume={2},
  number={1},
  pages={015006},
  year={2017},
  publisher={IOP Publishing}
}

@article{beigi2011quantum,
  title={The quantum double model with boundary: condensations and symmetries},
  author={Beigi, Salman and Shor, Peter W and Whalen, Daniel},
  journal={Communications in mathematical physics},
  volume={306},
  pages={663--694},
  year={2011},
  publisher={Springer}
}

@article{lu2009demonstrating,
  title={Demonstrating anyonic fractional statistics with a six-qubit quantum simulator},
  author={Lu, Chao-Yang and Gao, Wei-Bo and G{\"u}hne, Otfried and Zhou, Xiao-Qi and Chen, Zeng-Bing and Pan, Jian-Wei},
  journal={Physical review letters},
  volume={102},
  number={3},
  pages={030502},
  year={2009},
  publisher={APS}
}

@article{pachos2009revealing,
  title={Revealing anyonic features in a toric code quantum simulation},
  author={Pachos, JK and Wieczorek, Witlef and Schmid, Christian and Kiesel, Nikolai and Pohlner, Reinhold and Weinfurter, Harald},
  journal={New Journal of Physics},
  volume={11},
  number={8},
  pages={083010},
  year={2009},
  publisher={IOP Publishing}
}

@article{cui2015universal,
  title={Universal quantum computation with weakly integral anyons},
  author={Cui, Shawn X and Hong, Seung-Moon and Wang, Zhenghan},
  journal={Quantum Information Processing},
  volume={14},
  pages={2687--2727},
  year={2015},
  publisher={Springer}
}

@article{wilczek1982quantum,
  title={Quantum mechanics of fractional-spin particles},
  author={Wilczek, Frank},
  journal={Physical review letters},
  volume={49},
  number={14},
  pages={957},
  year={1982},
  publisher={APS}
}

@article{xu2023digital,
  title={Digital simulation of projective non-Abelian anyons with 68 superconducting qubits},
  author={Xu, Shibo and Sun, Zheng-Zhi and Wang, Ke and Xiang, Liang and Bao, Zehang and Zhu, Zitian and Shen, Fanhao and Song, Zixuan and Zhang, Pengfei and Ren, Wenhui and others},
  journal={Chinese Physics Letters},
  volume={40},
  number={6},
  pages={060301},
  year={2023},
  publisher={IOP Publishing}
}

@article{fiedler2015haag,
  title={Haag duality for Kitaev’s quantum double model for abelian groups},
  author={Fiedler, Leander and Naaijkens, Pieter},
  journal={Reviews in Mathematical Physics},
  volume={27},
  number={09},
  pages={1550021},
  year={2015},
  publisher={World Scientific}
}

@article{landon2013local,
  title={Local topological order inhibits thermal stability in 2D},
  author={Landon-Cardinal, Olivier and Poulin, David},
  journal={Physical review letters},
  volume={110},
  number={9},
  pages={090502},
  year={2013},
  publisher={APS}
}

@article{dennis2002topological,
  title={Topological quantum memory},
  author={Dennis, Eric and Kitaev, Alexei and Landahl, Andrew and Preskill, John},
  journal={Journal of Mathematical Physics},
  volume={43},
  number={9},
  pages={4452--4505},
  year={2002},
  publisher={American Institute of Physics}
}

@article{alicki2007statistical,
  title={A statistical mechanics view on Kitaev's proposal for quantum memories},
  author={Alicki, Robert and Fannes, M and Horodecki, Micha{\l}},
  journal={Journal of Physics A: Mathematical and Theoretical},
  volume={40},
  number={24},
  pages={6451},
  year={2007},
  publisher={IOP Publishing}
}

@article{alicki2009thermalization,
  title={On thermalization in Kitaev's 2D model},
  author={Alicki, Robert and Fannes, Mark and Horodecki, Michal},
  journal={Journal of Physics A: Mathematical and Theoretical},
  volume={42},
  number={6},
  pages={065303},
  year={2009},
  publisher={IOP Publishing}
}

@article{nussinov2008autocorrelations,
  title={Autocorrelations and thermal fragility of anyonic loops in topologically quantum ordered systems},
  author={Nussinov, Zohar and Ortiz, Gerardo},
  journal={Physical Review B—Condensed Matter and Materials Physics},
  volume={77},
  number={6},
  pages={064302},
  year={2008},
  publisher={APS}
}

@article{mourik2012signatures,
  title={Signatures of Majorana fermions in hybrid superconductor-semiconductor nanowire devices},
  author={Mourik, Vincent and Zuo, Kun and Frolov, Sergey M and Plissard, SR and Bakkers, Erik PAM and Kouwenhoven, Leo P},
  journal={Science},
  volume={336},
  number={6084},
  pages={1003--1007},
  year={2012},
  publisher={American Association for the Advancement of Science}
}

@article{bravyi2011short,
  title={A short proof of stability of topological order under local perturbations},
  author={Bravyi, Sergey and Hastings, Matthew B},
  journal={Communications in mathematical physics},
  volume={307},
  pages={609--627},
  year={2011},
  publisher={Springer}
}

@article{bravyi2010topological,
  title={Topological quantum order: stability under local perturbations},
  author={Bravyi, Sergey and Hastings, Matthew B and Michalakis, Spyridon},
  journal={Journal of mathematical physics},
  volume={51},
  number={9},
  year={2010},
  publisher={AIP Publishing}
}

@incollection{naaijkens2015kitaev,
  title={Kitaev’s quantum double model from a local quantum physics point of view},
  author={Naaijkens, Pieter},
  booktitle={Advances in algebraic quantum field theory},
  pages={365--395},
  year={2015},
  publisher={Springer}
}

@article{bonderson2009measurement,
  title={Measurement-only topological quantum computation via anyonic interferometry},
  author={Bonderson, Parsa and Freedman, Michael and Nayak, Chetan},
  journal={Annals of Physics},
  volume={324},
  number={4},
  pages={787--826},
  year={2009},
  publisher={Elsevier}
}

@article{bonderson2008measurement,
  title={Measurement-only topological quantum computation},
  author={Bonderson, Parsa and Freedman, Michael and Nayak, Chetan},
  journal={Physical review letters},
  volume={101},
  number={1},
  pages={010501},
  year={2008},
  publisher={APS}
}

@article{wootton2012quantum,
  title={Quantum memories and error correction},
  author={Wootton, James R},
  journal={Journal of Modern Optics},
  volume={59},
  number={20},
  pages={1717--1738},
  year={2012},
  publisher={Taylor \& Francis}
}

@inproceedings{wootton2009universal,
  title={Universal quantum computation with a non-Abelian topological memory},
  author={Wootton, James R and Lahtinen, Ville and Pachos, Jiannis K},
  booktitle={Theory of Quantum Computation, Communication, and Cryptography: 4th Workshop, TQC 2009, Waterloo, Canada, May 11-13, 2009, Revised Selected Papers 4},
  pages={56--65},
  year={2009},
  organization={Springer}
}

@article{wang2009threshold,
  title={Threshold error rates for the toric and surface codes},
  author={Wang, David S and Fowler, Austin G and Stephens, Ashley M and Hollenberg, Lloyd CL},
  journal={arXiv preprint arXiv:0905.0531},
  year={2009}
}

@article{iblisdir2009scaling,
  title={Scaling law for topologically ordered systems at finite temperature},
  author={Iblisdir, Sofyan and P{\'e}rez-Garc{\'\i}a, David and Aguado, Miguel and Pachos, J},
  journal={Physical Review B—Condensed Matter and Materials Physics},
  volume={79},
  number={13},
  pages={134303},
  year={2009},
  publisher={APS}
}

@article{kay2008quantum,
  title={Quantum self-correcting stabilizer codes},
  author={Kay, Alastair and Colbeck, Roger},
  journal={arXiv preprint arXiv:0810.3557},
  year={2008}
}

@article{prem2017glassy,
  title={Glassy quantum dynamics in translation invariant fracton models},
  author={Prem, Abhinav and Haah, Jeongwan and Nandkishore, Rahul},
  journal={Physical Review B},
  volume={95},
  number={15},
  pages={155133},
  year={2017},
  publisher={APS}
}

@article{terhal2015quantum,
  title={Quantum error correction for quantum memories},
  author={Terhal, Barbara M},
  journal={Reviews of Modern Physics},
  volume={87},
  number={2},
  pages={307--346},
  year={2015},
  publisher={APS}
}

@article{roberts2020symmetry,
  title={Symmetry-protected self-correcting quantum memories},
  author={Roberts, Sam and Bartlett, Stephen D},
  journal={Physical Review X},
  volume={10},
  number={3},
  pages={031041},
  year={2020},
  publisher={APS}
}

@article{bravyi2013quantum,
  title={Quantum self-correction in the 3d cubic code model},
  author={Bravyi, Sergey and Haah, Jeongwan},
  journal={Physical review letters},
  volume={111},
  number={20},
  pages={200501},
  year={2013},
  publisher={APS}
}

@article{yoshida2011feasibility,
  title={Feasibility of self-correcting quantum memory and thermal stability of topological order},
  author={Yoshida, Beni},
  journal={Annals of Physics},
  volume={326},
  number={10},
  pages={2566--2633},
  year={2011},
  publisher={Elsevier}
}

@article{shen2022fracton,
  title={Fracton topological order at finite temperature},
  author={Shen, Xiaoyang and Wu, Zhengzhi and Li, Linhao and Qin, Zhehan and Yao, Hong},
  journal={Physical Review Research},
  volume={4},
  number={3},
  pages={L032008},
  year={2022},
  publisher={APS}
}

@article{chamon2005quantum,
  title={Quantum Glassiness in Strongly Correlated Clean Systems: An Example of Topological Overprotection},
  author={Chamon, Claudio},
  journal={Physical review letters},
  volume={94},
  number={4},
  pages={040402},
  year={2005},
  publisher={APS}
}

@article{iqbal2024non,
  title={Non-Abelian topological order and anyons on a trapped-ion processor},
  author={Iqbal, Mohsin and Tantivasadakarn, Nathanan and Verresen, Ruben and Campbell, Sara L and Dreiling, Joan M and Figgatt, Caroline and Gaebler, John P and Johansen, Jacob and Mills, Michael and Moses, Steven A and others},
  journal={Nature},
  volume={626},
  number={7999},
  pages={505--511},
  year={2024},
  publisher={Nature Publishing Group UK London}
}

@article{duclos2010fast,
  title={Fast decoders for topological quantum codes},
  author={Duclos-Cianci, Guillaume and Poulin, David},
  journal={Physical review letters},
  volume={104},
  number={5},
  pages={050504},
  year={2010},
  publisher={APS}
}

@article{wootton2012high,
  title={High threshold error correction for the surface code},
  author={Wootton, James R and Loss, Daniel},
  journal={Physical review letters},
  volume={109},
  number={16},
  pages={160503},
  year={2012},
  publisher={APS}
}

@article{wootton2013topological,
  title={Topological phases and self-correcting memories in interacting anyon systems},
  author={Wootton, James R},
  journal={Physical Review A—Atomic, Molecular, and Optical Physics},
  volume={88},
  number={6},
  pages={062312},
  year={2013},
  publisher={APS}
}

@article{leinaas1977theory,
  title={On the theory of identical particles},
  author={Leinaas, JM and Myrheim, J},
  journal={Il nuovo cimento},
  volume={37},
  pages={132},
  year={1977}
}

@article{bravyi2005universal,
  title={Universal quantum computation with ideal Clifford gates and noisy ancillas},
  author={Bravyi, Sergey and Kitaev, Alexei},
  journal={Physical Review A—Atomic, Molecular, and Optical Physics},
  volume={71},
  number={2},
  pages={022316},
  year={2005},
  publisher={APS}
}

@article{cong2017universal,
  title={Universal quantum computation with gapped boundaries},
  author={Cong, Iris and Cheng, Meng and Wang, Zhenghan},
  journal={Physical Review Letters},
  volume={119},
  number={17},
  pages={170504},
  year={2017},
  publisher={APS}
}

@article{Siehler_2003,
   title={Near-group categories},
   volume={3},
   ISSN={1472-2747},
   url={http://dx.doi.org/10.2140/agt.2003.3.719},
   DOI={10.2140/agt.2003.3.719},
   number={2},
   journal={Algebraic \&; Geometric Topology},
   publisher={Mathematical Sciences Publishers},
   author={Siehler, Jacob},
   year={2003},
   month=aug, pages={719–775} }

@article{mochon2003anyons,
  title={Anyons from nonsolvable finite groups are sufficient for universal quantum computation},
  author={Mochon, Carlos},
  journal={Physical Review A},
  volume={67},
  number={2},
  pages={022315},
  year={2003},
  publisher={APS}
}

@article{wootton2011universal,
  title={Universal quantum computation with abelian anyon models},
  author={Wootton, James R and Pachos, Jiannis K},
  journal={Electronic Notes in Theoretical Computer Science},
  volume={270},
  number={2},
  pages={209--218},
  year={2011},
  publisher={Elsevier}
}

@article{bravyi2012magic,
  title={Magic-state distillation with low overhead},
  author={Bravyi, Sergey and Haah, Jeongwan},
  journal={Physical Review A—Atomic, Molecular, and Optical Physics},
  volume={86},
  number={5},
  pages={052329},
  year={2012},
  publisher={APS}
}

@article{howard2017application,
  title={Application of a resource theory for magic states to fault-tolerant quantum computing},
  author={Howard, Mark and Campbell, Earl},
  journal={Physical review letters},
  volume={118},
  number={9},
  pages={090501},
  year={2017},
  publisher={APS}
}

@article{aaronson2004improved,
  title={Improved simulation of stabilizer circuits},
  author={Aaronson, Scott and Gottesman, Daniel},
  journal={Physical Review A—Atomic, Molecular, and Optical Physics},
  volume={70},
  number={5},
  pages={052328},
  year={2004},
  publisher={APS}
}

@article{gottesman1998heisenberg,
  title={The Heisenberg representation of quantum computers},
  author={Gottesman, Daniel},
  journal={arXiv preprint quant-ph/9807006},
  year={1998}
}

@article{bombin2006topological,
  title={Topological quantum distillation},
  author={Bombin, Hector and Martin-Delgado, Miguel Angel},
  journal={Physical review letters},
  volume={97},
  number={18},
  pages={180501},
  year={2006},
  publisher={APS}
}

@book{lahtinen2010interacting,
  title={Interacting non-Abelian anyons in an exactly solvable lattice model},
  author={Lahtinen, Ville Tapani},
  year={2010},
  publisher={University of Leeds}
}

@article{chen2025universal,
  title={A universal circuit set using the S 3 quantum double},
  author={Chen, Liyuan and Ren, Yuanjie and Fan, Ruihua and Jaffe, Arthur},
  journal={npj Quantum Information},
  volume={11},
  number={1},
  pages={112},
  year={2025},
  publisher={Nature Publishing Group UK London}
}

@article{hormozi2007topological,
  title={Topological quantum compiling},
  author={Hormozi, Layla and Zikos, Georgios and Bonesteel, Nicholas E and Simon, Steven H},
  journal={Physical Review B—Condensed Matter and Materials Physics},
  volume={75},
  number={16},
  pages={165310},
  year={2007},
  publisher={APS}
}

@article{field2018introduction,
  title={Introduction to topological quantum computation with non-Abelian anyons},
  author={Field, Bernard and Simula, Tapio},
  journal={Quantum Science and Technology},
  volume={3},
  number={4},
  pages={045004},
  year={2018},
  publisher={IOP Publishing}
}

@inproceedings{aharonov2006polynomial,
  title={A polynomial quantum algorithm for approximating the Jones polynomial},
  author={Aharonov, Dorit and Jones, Vaughan and Landau, Zeph},
  booktitle={Proceedings of the thirty-eighth annual ACM symposium on Theory of computing},
  pages={427--436},
  year={2006}
}

@article{lo2025universal,
  title={Universal Quantum Computation with the $ S\_3 $ Quantum Double: A Pedagogical Exposition},
  author={Lo, Chiu Fan Bowen and Lyons, Anasuya and Verresen, Ruben and Vishwanath, Ashvin and Tantivasadakarn, Nathanan},
  journal={arXiv preprint arXiv:2502.14974},
  year={2025}
}

@book{dummit2004abstract,
  title={Abstract algebra},
  author={Dummit, David Steven and Foote, Richard M and others},
  volume={3},
  year={2004},
  publisher={Wiley Hoboken}
}

@article{grier2022classification,
  title={The classification of Clifford gates over qubits},
  author={Grier, Daniel and Schaeffer, Luke},
  journal={Quantum},
  volume={6},
  pages={734},
  year={2022},
  publisher={Verein zur F{\"o}rderung des Open Access Publizierens in den Quantenwissenschaften}
}

@article{buerschaper2009mapping,
  title={Mapping Kitaev’s quantum double lattice models to Levin and Wen’s string-net models},
  author={Buerschaper, Oliver and Aguado, Miguel},
  journal={Physical Review B—Condensed Matter and Materials Physics},
  volume={80},
  number={15},
  pages={155136},
  year={2009},
  publisher={APS}
}

@article{tounsi2023systematic,
  title={Systematic Computation of Braid Generator Matrix in Topological Quantum Computing},
  author={Tounsi, Abdellah and Belaloui, Nacer Eddine and Louamri, Mohamed Messaoud and Mimoun, Amani and Benslama, Achour and Rouabah, Mohamed Taha},
  journal={arXiv preprint arXiv:2307.01892},
  year={2023}
}

@article{tarabunga2023many,
  title={Many-body magic via pauli-markov chains—from criticality to gauge theories},
  author={Tarabunga, Poetri Sonya and Tirrito, Emanuele and Chanda, Titas and Dalmonte, Marcello},
  journal={PRX Quantum},
  volume={4},
  number={4},
  pages={040317},
  year={2023},
  publisher={APS}
}

@article{howard2014contextuality,
  title={Contextuality supplies the ‘magic’for quantum computation},
  author={Howard, Mark and Wallman, Joel and Veitch, Victor and Emerson, Joseph},
  journal={Nature},
  volume={510},
  number={7505},
  pages={351--355},
  year={2014},
  publisher={Nature Publishing Group UK London}
}

@article{knill2005quantum,
  title={Quantum computing with realistically noisy devices},
  author={Knill, Emanuel},
  journal={Nature},
  volume={434},
  number={7029},
  pages={39--44},
  year={2005},
  publisher={Nature Publishing Group UK London}
}

@article{campbell2012magic,
  title={Magic-state distillation in all prime dimensions using quantum reed-muller codes},
  author={Campbell, Earl T and Anwar, Hussain and Browne, Dan E},
  journal={Physical Review X},
  volume={2},
  number={4},
  pages={041021},
  year={2012},
  publisher={APS}
}

@article{anwar2012qutrit,
  title={Qutrit magic state distillation},
  author={Anwar, Hussain and Campbell, Earl T and Browne, Dan E},
  journal={New Journal of Physics},
  volume={14},
  number={6},
  pages={063006},
  year={2012},
  publisher={IOP Publishing}
}

@article{liu2023magic,
  title={Magic state distillation and cost analysis in fault-tolerant universal quantum computation},
  author={Liu, Yiting and Ma, Zhi and Luo, Lan and Du, Chao and Fei, Yangyang and Wang, Hong and Duan, Qianheng and Yang, Jing},
  journal={Quantum Science and Technology},
  volume={8},
  number={4},
  pages={043001},
  year={2023},
  publisher={IOP Publishing}
}

@inproceedings{campbell2009structure,
  title={On the structure of protocols for magic state distillation},
  author={Campbell, Earl T and Browne, Dan E},
  booktitle={Workshop on Quantum Computation, Communication, and Cryptography},
  pages={20--32},
  year={2009},
  organization={Springer}
}

@article{huang2025generating,
  title={Generating logical magic states with the aid of non-Abelian topological order},
  author={Huang, Sheng-Jie and Chen, Yanzhu},
  journal={arXiv preprint arXiv:2502.00998},
  year={2025}
}

@article{kim2024magic,
  title={Magic State Injection on IBM Quantum Processors Above the Distillation Threshold},
  author={Kim, Younghun and Sevior, Martin and Usman, Muhammad},
  journal={arXiv preprint arXiv:2412.01446},
  year={2024}
}

\begin{appendix}

\section{Quantum simulations vs actual physical realisations of anyons}
\label{sec:sims}

A natural question is how useful anyonic quantum simulations are, especially when considering minimal system sizes. To address this question we will consider various aspects of topological systems. A first desirable characteristic in topological quantum computation is the dynamical stability obtained by activating a Hamiltonian. It is known that the generated energy gap of topological Hamiltonians keeps the ground state resilient to erroneous perturbations \cite{bravyi2010topological, bravyi2011short}, which was the central characteristic that motivated the application of anyons to quantum computation \cite{kitaev2003fault}. Nevertheless, current quantum technology platforms have managed to minimise such perturbations to unprecedented levels making such a benefit against coherent errors rather redundant. The source of errors that impedes the realisation of quantum computers are incoherent errors, such as the presence of a finite temperature \cite{landon2013local, alicki2007statistical, alicki2009thermalization, nussinov2008autocorrelations, iblisdir2009scaling}. Typical anyonic systems are equally exposed to such errors as other quantum computation platforms, with fractons offering a promising direction for resolving this problem \cite{yoshida2011feasibility, shen2022fracton, chamon2005quantum, prem2017glassy}. Hence, the requirement of a Hamiltonian can be relaxed without diminishing the applicability of the anyonic simulations.

Beyond the Hamiltonian protection, topological states are equivalent to quantum error correcting codes (QECCs) with high error thresholds, minimal resource requirements and an intuitive syndrome measurement procedure that stems from the 2D geometry of the topological systems \cite{dennis2002topological, bacon2008stability}. In this dictionary, the logical space of the QECC is identified with the Hilbert space of the anyons. Hence, realisation and manipulations of anyonic states with quantum simulators is a step towards building efficient quantum error correction in the laboratory. Such QECCs offer the ability to correct a sufficiently low level of incoherent errors, although, unlike the Hamiltonian protection against coherent errors, they need active monitoring and correction \cite{wootton2009universal, wang2009threshold, kay2008quantum}. 

Another issue faced with quantum simulations is that their quantum degrees of freedom that encode the topological states do not necessarily possess a two-dimensional configuration. It should be noted that  the exact position e.g. of the links or vertices of the quantum double lattice, is not important for realising the desired anyonic properties. Topological invariance means the quantum properties of the system do not depend on a metric and hence geometric characteristics are not relevant.  Nevertheless, we do not want to lose the locality of the anyons, which stems from the locality of the plaquette and vertex operators of the quantum double model \cite{kitaev2003fault}. Indeed, as anyons are quasiparticles, they are expected to be localised in a finite region of the system. To preserve locality of the quantum double model we need to keep the information of the order of the qudits, e.g. by enumerating them, and we need to know their connectivity. This consideration also reflects on the locality of the syndrome measurements when the states are interpreted as a QECC, which is a desirable characteristic \cite{dennis2002topological, wootton2012quantum}.

We strongly believe that quantum simulation of small anyonic systems can became the seed for several future developments. The minimal system proposed here includes all the elements necessary to create the ecosystem of anyons and offers a proof-of-principle demonstration of anyonic properties. These few-qudit manipulations can be directly extended to larger systems when more qudits are available. Such an effort will allow the extension of our scheme to larger distance QECC and thus enhances the fault-tolerance of the system~\cite{wootton2014error}. Furthermore, implementing the anyonic protocols described here with larger systems benchmarks the available quantum technology towards various applications. Indeed, quantum double models can be directly connected to the fundamental physics of anyons \cite{wilczek1982quantum}, lattice gauge theories \cite{fiedler2015haag, naaijkens2015kitaev}, quantum error correction \cite{duclos2010fast, wootton2012high, wootton2013topological} and quantum computation \cite{pachos2012introduction, simon2023topological} thus providing a wide spectrum of applications.

\section{The Quantum Double Model $\mathbf{D}(\mathbf{G})$}
\label{sec:DG}

Here we provide an overview of the quantum double construction as first introduced in \cite{kitaev2003fault}. We start by outlining how the group $\mathbf{G}$ defines the algebraic structure of the quantum double $\mathbf{D}(\mathbf{G})$, dictating the anyons present and their associated fusion and braiding relations. We will then go on to present the Kitaev quantum double model construction, in which these anyons appear as point-like excitations on a lattice of spins.

\subsection{Anyons in the Quantum Double Model}
\label{sec:DGanyons}

For some finite group $\mathbf{G}$, the mathematical structure of the associated anyonic model is described by the Drinfield double of the group. The different types of anyons are in one-to-one correspondence with the irreducible representations (irreps) of this operator algebra, and may therefore be uniquely specified by $\left(\mathcal{C}_{g}, \Gamma^{\mathcal{N}_{g}}_{R(\mathcal{N}_{g})} \right)$ where $\mathcal{C}_{g}$ is the conjugacy class
denoting an anyons magnetic flux
\begin{equation}
    \mathcal{C}_{g} = \{hgh^{-1} \vert h\in \mathbf{G}\},
    \label{eq:Cg}
\end{equation}
$\mathcal{N}_{g}$ is the normalizer of a group measuring the anyon’s electric charge
\begin{equation}
    \mathcal{N}_{g} = \{gh=hg \vert h\in \mathbf{G}\},
\end{equation}
and $R(\mathcal{N}_{g})$ represents a unitary irrep of an element of the normaliser. Anyons that possess trivial charge (flux) are labelled fluxons (chargeons) and those with both non-trivial charge and flux are labelled dyons. We note that having trivial flux means that chargeons correspond to the conjugacy class $\mathcal{C}_{e} = \{e\}$, hence their normaliser will always be the group itself, $\mathcal{N}_{e} = \mathbf{G}$. 

For any anyon labelled $\left(\mathcal{C}_{g}, \Gamma^{\mathcal{N}_{g}}_{R(\mathcal{N}_{g})} \right)$ its quantum dimensionality is given by
\begin{equation}
    d_{\left(\mathcal{C}_{g}, \Gamma^{\mathcal{N}_{g}}_{R(\mathcal{N}_{g})} \right)} = |\mathcal{C}_{g}||\Gamma^{\mathcal{N}_{g}}_{R(\mathcal{N}_{g})}|,
\end{equation}
Abelian anyons necessarily have $d=1$ while all anyons with $d>1$ are non-Abelian.
The total quantum dimension, $\mathcal{D}$, of the model can be found from the sum of the quantum dimensions over all anyon types
\begin{equation}
    \mathcal{D}^{2} = \sum_{\text{anyons k}} d_{k}^{2},
\end{equation}
and is related to the cardinality of the chosen group
\begin{equation}
    \mathcal{D}^{2}=|\mathbf{G}|^{2}.
\end{equation}

\subsection{Ribbon Operators}
\label{sec:DGribbons}

\begin{figure}[!t]
    \centering
    \includegraphics[width=0.45\textwidth]{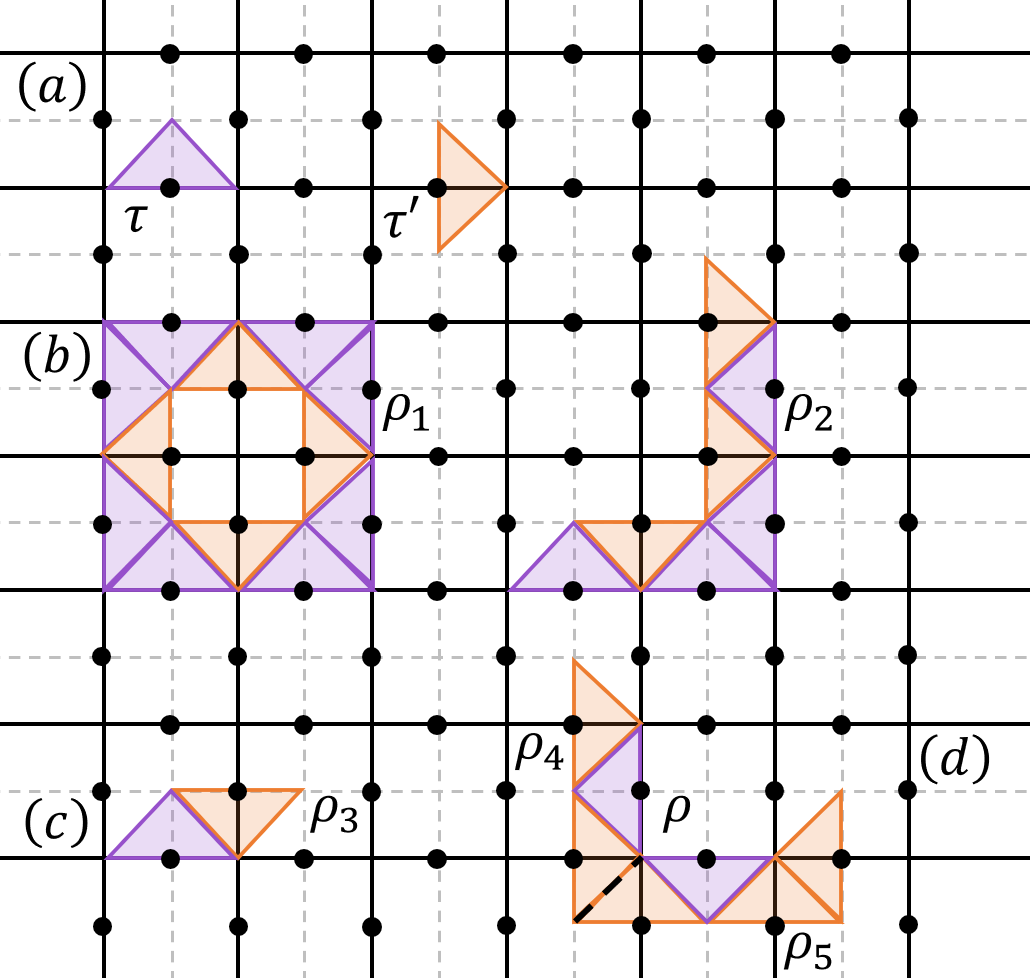}
    \caption{The square lattice of the quantum double model with various triangle and ribbon operators. (a) The direct $\tau$ and dual $\tau'$ triangles formed from the $L^g$ and $T^h$ operators, respectively. (b) Closed loop, $\rho_1$, and open-ended, $\rho_2$, ribbon operators formed from combinations of triangle operators. (c) A ribbon operator with one dual and one direct triangle. (d) Two ribbons $\rho_4$ and $\rho_5$ that have been `glued' together (shown by the dashed line on the connections of the triangles) to form a single extended ribbon $\rho$.}
    \label{fig:Lattice_ribbons}
\end{figure}

Having outlined the abstract formalism for the classification of anyons in the quantum double model, we will now provide more detail on the ribbon operators with which they may be coherently created and transported on a two-dimensional lattice as shown in Figure \ref{fig:Lattice}. 

For the quantum double $\mathbf{D(\mathbf{G})}$, an oriented lattice is constructed with a spin-like Hilbert space, $\mathbf{H}$, positioned on each link with orthonormal basis $\{ \ket{g}: g\in\mathbf{G}\}$, such that $\text{dim}(\mathbf{H}) = |\mathbf{G}|$. Four linear operators are introduced with action $L^{g}_{+}\ket{z}=\ket{gz}, L^{g}_{-}\ket{z}=\ket{zg^{-1}}, T^{h}_{+}\ket{z}=\delta_{h,z}\ket{z}, T^{h}_{-}\ket{z}=\delta_{h^{-1},z}\ket{z}$, for some $z,g,h\in\mathbf{G}$. The action of single-qudit operators $L^{g}_{\pm}$ and $T^{h}_{\pm}$ may be represented with dual and direct triangles respectively as shown in Fig.~\ref{fig:Lattice_ribbons}(a) (see Equations \eqref{eq:L} and \eqref{eq:T} for associated labelling convention). From these triangles, longer strips may be constructed as illustrated in Fig.~\ref{fig:Lattice_ribbons}(b).
Any such strip is a valid ribbon as long as their direct and dual paths do not self-cross \cite{bombin2008family}. For each ribbon $\rho$ we introduce a set of operators $\{F^{h,g}_{\rho}\}$ with $h,g \in \mathbf{G}$, called ribbon operators. These operators act on all qudits contained in $\rho$, however their action is only non-trivial on the start and end sites (i.e. the ribbon operators commute with $A(v)$ and $B^{e}(p)$ for all sites other than the end sites). In this way, stabiliser violations corresponding to excitations of the system only arise at the endpoints of the ribbon. 

In the simplest case, a ribbon may be constructed from a single triangle.
If $\tau$ is a dual triangle and $\tau^{\prime}$ a direct triangle, we set
\begin{equation}
    F^{h,g}_{\tau} := \delta_{1,g}L^{h}_{\tau}, \hspace{1cm} F^{h,g}_{\tau^{\prime}} := T^{g}_{\tau^{\prime}}.
    \label{eq:tri}
\end{equation}
These ribbons have minimal length $l=1$. 

A ribbon is described as ‘proper' if it is composed of at least one dual and one direct triangle. The smallest proper ribbon is thus composed of exactly one direct and one dual triangle as shown in Fig.~\ref{fig:Lattice_ribbons}(c). The associated ribbon operator is defined as the product of the ribbon operators on each triangle,
\begin{equation}
    F^{h,g}_{\rho} = L^{h}_{\tau_{\text{dual}}}T^{g}_{\tau_{\text{dir}}}
    \label{eq:proper}
\end{equation}

For all general ribbons with $l>1$, we may let $\rho = \rho_{1}\rho_{2}$ and recursively define a ‘gluing' procedure with the following
\begin{equation}
    F^{h,g}_{\rho} := \sum_{k\in\mathbf{G}} F_{\rho_{1}}^{h,k} F_{\rho_{2}}^{\bar{k}hk, \bar{k}g}
    \label{eq:glue}
\end{equation}
where the first site of $\rho_{2}$ is always the same as the last site of $\rho_{1}$ (see Fig.~\ref{fig:Lattice_ribbons}(d)).

These operators, $F^{h,g}_{\rho}$, allow for the construction of ribbon operators that give rise to $X$ anyons at the endpoints of the path $\rho$. The construction of these is non-trivial as, unlike 
with the toric code \cite{kitaev2003fault}, excitations in the form of spin rotations do not necessarily correspond to the creation of anyons \cite{komar2017anyons}.

For any Abelian anyon, ribbon operators giving rise to $X$ anyons at the endpoints of the path $\rho$ are given by \cite{fiedler2015haag}
\begin{equation}
    F_{\rho}^{\chi,c} = \sum_{g\in \mathbf{G}} \bar{\chi}(g)F^{\bar{c},g}_{\rho}
    \label{eq:Fa}
\end{equation}
where $c$ is an element of $\mathbf{G}$ and $\chi$ is a character representation of $\mathbf{G}$.

For non-Abelian anyons we must introduce the non-Abelian basis transformation \cite{bombin2008family}
\begin{equation}
    F^{R\mathcal{C};\textbf{uv}}:= \frac{n_{R}}{|\mathcal{N}_{C}|} \sum_{n\in\mathcal{N}_{\mathcal{C}}} \bar{\Gamma}_{R}^{j,j^{\prime}}(n) F^{\bar{c}_{i},q_{i}nq_{i^{\prime}}}
    \label{eq:FnonA}
\end{equation}
with each anyon having a unique conjugacy class $\mathcal{C}$ and $R$, the irrep of the normaliser corresponding to $\mathcal{C}$,
$\mathcal{N}_{\mathcal{C}}$ . The unitary matrices of the irrep are given as
$\Gamma_{R}(g)$, $g \in \mathbf{G}$ and the dimension of this is given by $n_{R}$. Each element of the normaliser and conjugacy class are given by $n_{i}$ and $c_{i}$ respectively. The representation $Q_{\mathcal{C}}$ is defined by $\mathbf{G}/\mathcal{N}_{\mathcal{C}}$ and each element is given by $q_{i}$. Finally, $\mathbf{u} = (i, j)$ and $\mathbf{v} = (i^{\prime}, j^{\prime})$ are additional degrees of freedom that may be traced out to give the final anyonic ribbon operator \cite{laubscher2019universal}. Examples of this construction for the Abelian $A$, $B$ and non-Abelian $G$ anyons of our chosen model are given in Appendix \ref{sec:S3}.


\section{Full Anyon Model $\mathbf{D}(\mathbf{S}_{3})$}
\label{sec:S3}

The $\mathbf{D}(\mathbf{S}_{3})$ quantum double model is based on the group structure of $\mathbf{S}_{3}$, which is isomorphic to the symmetry transformations of an equilateral triangle, given by
\begin{equation}
    \mathbf{S}_{3}=\{e,c,c^{2},t,tc,tc^{2} \},
\end{equation}
where $c$ is the generator for rotations following $c^{3}=e$,
and $t$ is the generator for reflections following $t^{2}=e$ with
$e$ as the identity element. The non-Abelian nature of this group is defined by the relation $ct = tc^{2}$. 

The anyons of the Drinfeld double of $\mathbf{S}_{3}$ are labeled by the conjugacy classes of $\mathbf{S}_{3}$ and the irreducible representations of the normalizers of these conjugacy classes. There are
three conjugacy classes of $S_{3}$:
\begin{equation}
    \begin{gathered}
    \{\mathcal{C}_{e}\} = \{e\}, \\
    \{\mathcal{C}_{c}\} = \{c,c^{2}\}, \\
    \{\mathcal{C}_{t}\} = \{t,tc,tc^{2}\},
    \end{gathered}
\end{equation}
with corresponding normalizers,
\begin{equation}
    \begin{gathered}
        \mathcal{N}_{e} = S_{3}, \\
        \mathcal{N}_{c}=\mathcal{N}_{c^{2}}=\{e,c,c^{2}\}\cong \mathbb{Z}_{3}, \\
        \mathcal{N}_{t} = \{e,t\}\cong \mathcal{N}_{tc} \cong \mathcal{N}_{tc^{2}} \cong \mathbb{Z}_{2}
    \end{gathered}
\end{equation}
where ‘$\cong$' denotes an isomorphism \cite{komar2017anyons}. The $\mathbf{D}(\mathbf{S}_{3})$ quantum double model therefore consists of eight types of anyons labelled $\{A, B, C, D, E, F, G, H\}$, where each type of anyon corresponds to one of the eight irreducible representations of $\mathbf{D}(\mathbf{S}_{3})$ as detailed in Table \ref{tab:ds3 anyons}. $A$ denotes the vacuum particle with both trivial flux and charge, such that $A\times Z = Z$ for $Z\in\{A,B,\dots,H\}$. 

\hspace{0.01cm}
\begin{table}[!ht]
    \centering
\begin{tabular}{ | c || c c c c c | }
    \hline
    Anyon & $\mathcal{C}_{g}$ & $\mathcal{N}_{g}$ & Irrep. & Q-dim. & Type \\ 
        [0.5ex]
     \hline
     \hline
    A & $\mathcal{C}_{e}$ & $S_{3}$ & $\Gamma_{1}^{S_{3}}$ & 1 & Vacuum\\
    \hline
    B & $\mathcal{C}_{e}$ & $S_{3}$ & $\Gamma_{-1}^{S_{3}}$ & 1 & Chargeon\\
    \hline
    C & $\mathcal{C}_{e}$ & $S_{3}$ & $\Gamma_{2}^{S_{3}}$ & 2 & Chargeon\\
    \hline
    \hline
    D & $\mathcal{C}_{t}$ & $\mathbb{Z}_{2}$ & $\Gamma_{1}^{\mathbb{Z}_{2}}$ & 3 & Fluxon\\
    \hline
    E & $\mathcal{C}_{t}$ & $\mathbb{Z}_{2}$ & $\Gamma_{-1}^{\mathbb{Z}_{2}}$ & 3 & Dyon\\
    \hline
    \hline
    F & $\mathcal{C}_{c}$ & $\mathbb{Z}_{3}$ & $\Gamma_{1}^{\mathbb{Z}_{3}}$ & 2 & Fluxon\\
    \hline
    G & $\mathcal{C}_{c}$ & $\mathbb{Z}_{3}$ & $\Gamma_{\omega}^{\mathbb{Z}_{3}}$ & 2 & Dyon\\
    \hline
    H & $\mathcal{C}_{c}$ & $\mathbb{Z}_{3}$ & $\Gamma_{\bar{\omega}}^{\mathbb{Z}_{3}}$ & 2 & Dyon \\
    \hline
\end{tabular}
    \caption{Anyons of $\mathbf{D}(S_{3})$ with their charge and flux labels, quantum dimensions and type.}
    \label{tab:ds3 anyons}
\end{table}

In this work, we restrict to the non-Abelian sub-group $\{A,B,G\}$ which is closed under fusion as seen in Equation \eqref{eq:fusionrel}. 
The irreps and character representations used in the construction of the ribbon operator creating pairs of these anyons as in \eqref{eq:Fa} and \eqref{eq:FnonA} are detailed in Table \ref{tab:ABG} \cite{goel2023unveiling}.\\

\begin{table}[!ht]
    \centering
\begin{tabular}{ | c | c || c c c c c c | }
    \hline
    Anyon & Irrep. & $e$ & $c$ & $c^{2}$ & $t$ & $tc$ & $tc^{2}$ \\ 
        [0.5ex]
     \hline
     \hline
    $A$ & $\Gamma_{1}^{S_{3}}$ & 1 & 1 & 1 & 1 & 1 & 1\\
    \hline
    $B$ & $\Gamma_{-1}^{S_{3}}$ & 1  & 1 & 1 & -1 & -1 & -1\\
    \hline
    $G$ & $\Gamma_{\omega}^{\mathbb{Z}_{3}}$ & 1 & $\omega$ & $\bar{\omega}$ & 0 & 0 & 0\\
    \hline
\end{tabular}
    \caption{The irreducible representations of the submodel $\{A,B,G\}$ and the traces of their corresponding character irreducible representations.}
    \label{tab:ABG}
\end{table}

Both Abelian anyons in our sub-model, $A:(\mathcal{C}_{e}=\{e\},\Gamma^{\mathbf{S}_{3}}_{1})$ and $B:(\mathcal{C}_{e},\Gamma^{\mathbf{S}_{3}}_{-1})$ in the $\mathbf{D}(\mathbf{S}_{3})$ quantum double model correspond to the trivial conjugacy class $\mathcal{C}_{e}$ and are therefore chargeons residing on the vertices of the lattice. The simplest ribbon capable of producing a pair of such anyons is a single-qudit direct triangle $\rho=\tau$ as shown in Figure \ref{fig:Lattice}(a). The forms of Eq.\eqref{eq:FA} and \eqref{eq:FB} arise from the substitution of this information into Equation \eqref{eq:Fa} with the character representations of Table \ref{tab:ABG}. For example, for a pair of $B$ anyons
\begin{align}
    F^{B}_{\tau} &\equiv F^{\chi(\Gamma^{\mathbf{S}_{3}}_{-1}),e}_{\tau}, \\
    &= F^{e,e}_{\tau} + F^{e,c}_{\tau} + F^{e,c^{2}}_{\tau} - F^{e,t}_{\tau} - F^{e,tc}_{\tau} - F^{e,tc^{2}}_{\tau}, \\
    &= T^{e}_{+} + T^{c}_{+} + T^{c^{2}}_{+} - T^{t}_{+} - T^{tc}_{+} - T^{tc^{2}}_{+},
\end{align}
where we have applied the identity $F^{g,h}_{\tau}:= T^{h}_{\tau}$ for a direct triangle $\tau$ seen in Equation \eqref{eq:tri} and without loss of generality assumed that the triangle is facing along the oriented edge.

Similarly, for the non-Abelian anyons $G:(\mathcal{C}_{c},\Gamma^{\mathbf{Z}_{3}}_{\omega})$ with irreducible representation as given in Table \ref{tab:ABG}, Equation \eqref{eq:FnonA} yields
\begin{equation}
    F^{G}_{\rho} = F_{\rho}^{c,e}+\omega F_{\rho}^{c,c}+\bar{\omega}F_{\rho}^{c,c^{2}}+F_{\rho}^{c^{2},e}+\bar{\omega}F_{\rho}^{c^{2},c}+\omega F_{\rho}^{c^{2},c^{2}}.
\end{equation}
The $G$ anyon is dyonic, such that any ribbon $\rho$ along which this operator may be applied, must be composed of at least one dual and one direct triangle.
For ribbons $\rho_{1}$ and $\rho_{2}$ of the form $\rho=\tau_{\text{direct}}\tau_{\text{dual}}$, Equation \eqref{eq:proper} is used to find the specific ribbon operators $F^{G}_{\rho_{1}}$ and $F^{G}_{\rho_{2}}$ given in Equations \eqref{eq:rho_1} and \eqref{eq:rho_2} respectively.


\subsection{$R$ and $F$ Matrices}

As briefly introduced in Section \ref{sec:double_mod}, the sub-model $\{A,B,G\}$ has a non-trivial fusion as given in the fuson rules set out in \eqref{eq:fusionrel}, where the fusion multiplicity of the $G$ anyon signals its non-Abelian nature and subsequent potential for performing the non-trivial braiding key to the implementation of topological quantum computing schemes. Two $G$ anyons have a three-dimensional Hilbert space indexed by each of the three orthogonal fusion channels as given in equation \eqref{eq:fusionrel}. In demonstrating the non-triviality of the braiding of these anyons it is constructive to begin by introducing the $F$ and $R$ matrices, the basic operations for the manipulation of anyons. 

When two anyons labelled $a$ and $b$ undergo a counter-clockwise exchange, fusion commutativity $a\times b=b\times a$ dictates that their total charge $c$ remains unchanged. Assigning a distinct basis state to each configuration, the swapping of two particles on a line therefore induces an isomorphism between these states described by the braiding operator
\begin{equation}
R_{\circlearrowleft}\ket{a,b\rightarrow c} = R^{ab}_{c}\ket{b,a\rightarrow c}
    \label{eq:Rabc_app}
\end{equation}
where $R_{\circlearrowleft}$ corresponds to the anti-clockwise exchange of $a$ and $b$ anyons, as shown in Figure \ref{fig:R_and_F}(a). For fusion processes with a unique fusion channel this mapping corresponds to the possible acquisition of a simple phase factor, $R^{ab}_c$. For the non-Abelian fusion of two G anyons, the fusion multiplicity generates a $3\times3$ diagonal unitary matrix $R^{GG}$ with entries $(R^{GG} )_{i}^{i}\equiv R_{i}^{GG}$, the form of which may be analytically determined using a set of self-consistent relations known as the pentagon and hexagon relations \cite{pachos2012introduction}. The matrix $R^{GG}$ is found to take the form
\begin{equation}
    R^{GG} = \begin{pmatrix}
        \omega & 0 & 0 \\
        0 & -\omega & 0 \\
        0 & 0 & \bar{\omega}
    \end{pmatrix}, 
\end{equation}
with $\omega=e^{\frac{2\pi i}{3}}$ and $\bar{\omega}=\omega^{*}$.

When considering the fusion of three anyons $a,b,c$ to some fixed outcome $d$, the associated Hilbert space may be indexed by the possible intermediate fusion outcomes $a\times b\rightarrow i$ such that $i\times c\rightarrow d$ as shown on the left-hand side of Figure \ref{fig:R_and_F}(b). Each state in this Hilbert space compactly takes the form $\ket{(a,b),c \rightarrow i,c\rightarrow d}$. Fusion associativity $(a\times b)\times c=a\times(b\times c)$ dictates that this same Hilbert space can equivalently be spanned by the states as represented on the right-hand side of Figure \ref{fig:R_and_F}(b), for intermediary composite anyons $j$. Mapping between these two representations therefore requires the description of a basis transformation known as the fusion matrix $F_{abc}^{d}$. The action of this mapping may be described in the anyonic basis as
\begin{equation}
    \ket{(a,b),c\rightarrow i,c\rightarrow d} = \sum_{j} (F^{d}_{abc})^{i}_{j} \ket{a,(b,c)\rightarrow a,j\rightarrow d}.
\end{equation}

The fusion and braiding matrices $R^{GG}$ and $F^{G}_{GGG}$ provide a method of producing the operator of the $G$ anyonic braid group $b^G_2$, as shown in Eq.~\eqref{eq:b2}. 


In a recent work \cite{goel2023unveiling} it was demonstrated how the matrix $(R^{GG})^2$ can be reproduced with the application of a pair of single qutrit $G$ ribbons on a lattice. This minimal approach on a single qutrit could be efficiently simulated and verified using a quantum process tomography, thus providing the first step towards experimentally verifying the non-Abelianity of the $G$ anyons in $\mathbf{D}(\mathbf{S}_{3})$. Here, we build on this approach to introduce an alternative set of minimal operators with which we not only re-derive $(R^{GG})^2$, but also explicitly enact the swapping of the fusion order necessary to determine $F^{G}_{GGG}$. In this way we provide the first explicit verification of the non-commutativity of $(R^{GG})^2$ and $F^{G}_{GGG}$. Below we introduce the necessary components to realise the quantum double $\mathbf{D}(\mathbf{S}_{3})$ on a lattice. 

\subsection{Lattice representation of the $\mathbf{D}(\mathbf{S}_3)$ model}
\label{sec:LatticeRep}

We now present the lattice model that gives rise to the quantum double $\mathbf{D}(\mathbf{S}_{3})$ presented above. 
To this end, we introduce the two-dimensional square lattice with orientation as shown in Figure \ref{fig:Lattice}. On each link of the lattice a Hilbert space described by $d=6$ qudits is placed with orthonormal basis indexed by the group elements of $\mathbf{S}_{3}$.

Four linear operators are introduced with action $L^{g}_{+}\ket{z}=\ket{gz}, L^{g}_{-}\ket{z}=\ket{zg^{-1}}, T^{h}_{+}\ket{z}=\delta_{h,z}\ket{z}, T^{h}_{-}\ket{z}=\delta_{h^{-1},z}\ket{z}$, for some $z,g,h\in\mathbf{S}_{3}$, on any six-dimensional qudit. These operators are represented by dual and direct triangles respectively, as shown in Figure \ref{fig:Lattice}(a). 
From these single-qudit operators a set of mutually commuting vertex 
\begin{align}
    A(v) &= \frac{1}{6} \sum_{g\in\mathbf{S}_{3}} A^{g}(v), \nonumber \\
    &= \frac{1}{6} \sum_{g\in\mathbf{G}} L^{g}_{\pm}\otimes L^{g}_{\pm}\otimes L^{g}_{\pm}\otimes L^{g}_{\pm},
\end{align}
and plaquette operators
\begin{equation}
    B^{h}(p)=\sum_{h_{1}h_{2}h_{3}h_{4}=h} T^{h_{1}}_{\pm}\otimes T^{h_{2}}_{\pm}\otimes T^{h_{3}}_{\pm}\otimes T^{h_{4}}_{\pm},
\end{equation}
may be constructed from closed loops of triangles as shown in Fig.~\ref{fig:Lattice}(b). The choice of $L^{g}_{\pm}$ and $T^{h}_{\pm}$ in each operator are chosen in accordance with the following orientation convention
\begin{equation}
  L^{g}_{\pm} =
    \begin{cases}
      L^{g}_{+} & \text{if edge is oriented into vertex,}\\
      L^{g}_{-} & \text{if edge is oriented out of vertex,}
    \end{cases}   
    \label{eq:L}
\end{equation}
\begin{equation}
  T^{h}_{\pm}  =
    \begin{cases}
      T^{h}_{+} & \text{if edge is clockwise}\\
                 & \qquad \text{w.r.t. the plaquette,}\\
      T^{h}_{-} & \text{if edge is anti-clockwise} \\
                 & \qquad \text{w.r.t. the plaquette.}\\
    \end{cases}  
    \label{eq:T}
\end{equation}
such that the vertex and plaquette operators as shown in Fig.~\ref{fig:Lattice}(b) correspond to
$A^g(v) = L^{g}_{+}\otimes L^{g}_{-}\otimes L^{g}_{-}\otimes L^{g}_{+}$ and $B^{h}(p)=\sum T^{h_{1}}_{-}\otimes T^{h_{2}}_{-}\otimes T^{h_{3}}_{+}\otimes T^{h_{4}}_{+}$ respectively. Assigning such an orientation convention to the single qudit operators ensures that all operators $A(v)$ and $B(p)\equiv B^{e}(p)$ are mutually commuting as discussed in Appendix \ref{sec:comm}.

The Kitaev Hamiltonian \cite{kitaev2003fault}
\begin{equation}
    \mathcal{H} = -\sum_{v} A(v) -\sum_{p} B(p),
    \label{eq:App_H}
\end{equation}
defined on a lattice with open boundaries, produces a unique ground state $\ket{\zeta}$, corresponding to the anyonic vacuum, in which 
\begin{equation}
    A(v)\ket{\zeta}=\ket{\zeta}, \hspace{0.2cm} B(p)\ket{\zeta}=\ket{\zeta},
    \label{eq:App_gscond}
\end{equation} 
for all vertices, $v$, and plaquettes, $p$. Particle-like excitations of this model are indicated by violations of the conditions \eqref{eq:gscond}.

The $A$ and $B$ anyons are both Abelian with trivial flux, and can therefore be created and moved with string operators corresponding to paths on the direct lattice as seen in Appendix \ref{sec:DGribbons}. In the simplest case, pairs of these anyons can be created on neighbouring vertices with the single-qudit operators
\begin{align}
    F^{A}_{\tau} 
    &= T^{e}_{+} + T^{c}_{+}+T^{c^{2}}_{+} + T^{t}_{+} + T^{tc}_{+}+T^{tc^{2}}_{+} \nonumber\\
    &= \ket{e}\bra{e}+\ket{c}\bra{c} + \ket{c^{2}}\bra{c^{2}} \label{eq:FA} \\
    & \qquad \quad+ \ket{t}\bra{t} + \ket{tc}\bra{tc} +\ket{tc^{2}}\bra{tc^{2}} \nonumber \\
    &\equiv \mathbf{1}_{6} \nonumber
\end{align}
\begin{align}
    F^{B}_{\tau} 
    &= T^{e}_{+} + T^{c}_{+}+T^{c^{2}}_{+} - T^{t}_{+} - T^{tc}_{+}-T^{tc^{2}}_{+} \nonumber\\
    &= \ket{e}\bra{e}+\ket{c}\bra{c} + \ket{c^{2}}\bra{c^{2}} \label{eq:FB} \\
    & \qquad \quad- \ket{t}\bra{t} - \ket{tc}\bra{tc} - \ket{tc^{2}}\bra{tc^{2}} \nonumber \label{eq:FB}
\end{align}
represented as the direct triangles shown in Figure \ref{fig:Lattice}(a).

The $G$ anyons are dyons that have both non-trivial flux and charge. They therefore require the implementation of ribbon operators composed of dual and direct triangles for their manipulation. In the following we will consider two distinct two-qudit ribbon operators
\begin{equation}
    F^{G}_{\rho_{1}} = (T^{e}_{-} + \omega T^{c}_{-} + \bar{\omega} T^{c^{2}}_{-}) \otimes L^{c}_{-} + (T^{e}_{-} + \bar{\omega} T^{c}_{-} + \omega T^{c^{2}}_{-}) \otimes L^{c^{2}}_{-} \label{eq:rho_1}
\end{equation}
and
\begin{equation}
    F^{G}_{\rho_{2}} = L^{c}_{+}\otimes (T^{e}_{-} + \omega T^{c}_{-} + \bar{\omega} T^{c^{2}}_{-}) + L^{c^{2}}_+\otimes (T^{e}_{-} + \bar{\omega} T^{c}_{-} + \omega T^{c^{2}}_{-}) \label{eq:rho_2}
\end{equation}
each composed of one direct and one dual triangle, as illustrated in Figure \ref{fig:Lattice}(c).

Finally, we will introduce a set of local measurement operators that may measure a specific anyon type laying on the selected vertex. For this purpose we refer to the four-body charge projection operators as introduced in \cite{komar2017anyons}. Here, the charge projector associated to an irreducible representation $\Gamma$ of the chosen group $\mathbf{G}$ is given by
\begin{equation}
    A_{\Gamma}(v) = \frac{d_{\Gamma}}{|\mathbf{G}|} \sum_{g\in\mathbf{G}} \chi_{\Gamma}(g) A^{g}(v),
\end{equation}
where $d_{\Gamma}$ is the dimension of the irrep $\Gamma$ and $\chi_{\Gamma}(g)=Tr[\Gamma(g)]$ is the character of the group element $g$ in irrep $\Gamma$. For the three irreducible representations of $\mathbf{S}_{3}$ we thus have
\begin{equation}
    \begin{split}
    &  A_{\Gamma^{\mathbf{S}_3}_{1}}(v) = \frac{1}{6} \Big(A^e(v) + A^c(v) + A^{c^2}(v) + A^t(v) + A^{tc}(v) + A^{tc^2}(v)\Big),\\
    & A_{\Gamma^{\mathbf{S}_3}_{-1}}(v) = \frac{1}{6} \Big(A^e(v) + A^c(v) + A^{c^2}(v) - A^t(v) - A^{tc}(v) - A^{tc^2}(v)\Big), \\
    & A_{\Gamma^{\mathbf{S}_3}_{2}}(v) = \frac{1}{3} \Big(2A^{e}(v) - A^{c}(v)-A^{c^{2}}(v)\Big).
    \end{split}
    \label{eq:Avproj}
\end{equation}
These vertex projection operators form a set of orthogonal projective measurements such that
\begin{equation}
    \begin{split}
        \sum_{\Gamma\in\Gamma^{\mathbf{S}_{3}}} A_{\Gamma}(v) = \mathbf{1}, \hspace{2cm}\\
        A_{\Gamma}(v)A_{\Lambda}(v) = \delta_{\Gamma,\Lambda} A_{\Gamma}(v), \hspace{0.4cm} \Gamma,\Lambda\in \Gamma^{\mathbf{S}_{3}}\\
    \end{split}
    \label{eq:projorth}
\end{equation}
and are thus sufficient in uniquely distinguishing between the possible anyonic charges that lay on a chosen vertex $v$. 

Within the $\mathbf{D}(\mathbf{S}_{3})$ quantum double model the $A$ and $B$ anyons are both chargeons with charge component labelled by e one-dimensional irreps of $\mathbf{S}_{3}$ (see Table \ref{tab:ABG}). In this way the unique charge measurements of each anyon can be identified as $A^{A}(v)\equiv A_{\Gamma^{S_3}_{1}}(v)$ and $A^{B}(v)\equiv A_{\Gamma^{S_3}_{1}}(v)$. As this anyonic subgroup is closed under fusion, all remaining charge detection is therefore identified with that of the $G$ anyons, thus $A^{G}(v)\equiv A_{\Gamma^{S_3}_{2}}(v)$. In this way, the detection of each anyon type within our model is performed by a set of orthonormal charge projection operators with action as illustrated in Section \ref{sec:DG}.


\section{Magic state generation through braiding of $G$ anyons}
\label{sec:mbraid}


Consider the creation of two pairs of $G$ anyons from the vacuum as shown in Figure \ref{fig:b1b2}(a). The subsequent fusion of the pair $(G_{1},G_{2})$ has three possible outcomes $i=A,B,G$. As each of these anyons is its own antiparticle, the pair $(G_{3},G_{4})$ must fuse to the same composite anyon $i$ such that the total system stays within the vacuum superselection sector. One therefore obtains a three-dimensional fusion basis with states specified by the intermediate anyon $i$.
Alternatively, with the omission of the paths of trivial vacuum anyons, this system may be viewed as the fusion of three $G$ anyons to fixed outcome $G$ as shown in Figure \ref{fig:b1b2}(a). With this fixed order of fusion the three-dimensional basis may be explicitly spanned by the states $\{\ket{i}\}\equiv\{\ket{(G_{1},G_{2}),G_{3}\rightarrow i,G_{3}\rightarrow G_{4}}\}$ for $i=A,B,G$. 
As introduced in the previous chapter, anyonic associativity means that this fusion of three $G$ anyons to a fixed outcome may equivalently correspond to the basis  $\ket{j^{\prime}}=\ket{G_{1},(G_{2},G_{3})\rightarrow G_{1},j^{\prime}\rightarrow G_{4}}$, with the two bases related by the matrix $F^{G}_{GGG}$ through $\ket{i} = \sum_{j^{\prime}} (F^{G}_{GGG})^{i}_{j^{\prime}} \ket{j^{\prime}}$, and $\ket{j^{\prime}} = \sum_{i} (F^{G \hspace{0.2cm}-1}_{GGG})^{j^{\prime}}_{i} \ket{i}$.

The topologically distinct paths of $G_{1}, G_{2}$ and $G_{3}$ prior to fusion form the 3-strand braid group $\mathcal{B}_{3}$. This group has two generators $b_{1,G}$ and $b_{2,G}$ braiding pairs $(G_{1},G_{2})$ and $(G_{2},G_{3})$ respectively as shown.
When acting on a defined fusion space, the action of these braiding operations may be expressed as a product of $R$ and $F$ matrices. Consider for example $b_{1,G}\ket{i}$ as shown in Figure \ref{fig:b1b2}. The anyons $G_{1}$ and $G_{2}$ share a direct fusion channel such that we have simply
\begin{equation}
    b_{1,G}\ket{i} = R^{GG}\ket{i}.
\end{equation}
In contrast, in the basis $\{\ket{i}\}$ the anyons $G_{2}$ and $G_{3}$ do not fuse directly and thus the matrix $F^{G}_{GGG}$ must be used to form a description of the operation $b_{2,G}$ \cite{simon2023topological}.
Through the series of transformations illustrated in Figure \ref{fig:b2proof} one obtains
\begin{align}
    b_{23}\ket{x} &= b_{23} \left(\sum_{x^{\prime}} (F^{G}_{GGG})^{x}_{x^{\prime}} \ket{x^{\prime}}\right), \nonumber\\
    &= \sum_{x^{\prime}} R^{GG}_{x^{\prime}}(F^{G}_{GGG})^{x}_{x^{\prime}} \ket{x^{\prime}}, \nonumber\\
    &= \sum_{x^{\prime}} R^{GG}_{x^{\prime}}(F^{G}_{GGG})^{x}_{x^{\prime}} \left(\sum_{y} (F^{G}_{GGG})^{x^{\prime}}_{y}\ket{y}\right), \nonumber\\
    &= \sum_{y} \left(\sum_{x^{\prime}} (F^{G \hspace{0.1cm} -1}_{GGG})^{x^{\prime}}_{y} R^{GG}_{x^{\prime}}(F^{G}_{GGG})^{x}_{x^{\prime}}\right) \ket{y}.
\end{align}
Consider now the operator product $F^{G \hspace{0.1cm} -1}_{GGG}R^{GG}F^{G}_{GGG}$. The matrix elements of this operator take the form $(F^{G \hspace{0.1cm} -1}_{GGG}R^{GG}F^{G}_{GGG})^{y}_{x} = \sum_{n,m} (F^{G \hspace{0.1cm} -1}_{GGG})^{y}_{n} (R^{GG})^{n}_{m}(F^{G}_{GGG})^{m}_{x}$. For the diagonal $R$ matrix, $(R^{GG})^{n}_{m}=\delta_{n,m}R^{GG}_{n}$ such that this expression collapses to $(F^{G \hspace{0.1cm} -1}_{GGG}R^{GG}F^{G}_{GGG})^{y}_{x} = \sum_{n} (F^{G \hspace{0.1cm} -1}_{GGG})^{y}_{n} R^{GG}_{n}(F^{G}_{GGG})^{m}_{x}$. Finally, the Hermiticity of the real matrix $F^{G}_{GGG}$ implies both $F^{G \hspace{0.1cm} -1}_{GGG}= F^{G}_{GGG}$ and $(F^{G}_{GGG})^{n}_{m}=(F^{G}_{GGG})^{m}_{n}$ such that
\begin{equation}
    F^{G \hspace{0.1cm} -1}_{GGG}R^{GG}F^{G}_{GGG} =  \sum_{n} (F^{G \hspace{0.1cm} -1}_{GGG})^{n}_{y} R^{GG}_{n}(F^{G}_{GGG})^{x}_{n}, \nonumber
\end{equation}
and therefore
\begin{equation}
    b_{2,G}\ket{x} = F^{G \hspace{0.1cm} -1}_{GGG}R^{GG}F^{G}_{GGG}\ket{x}.
\end{equation}


Direct substitution of $R^{GG}$ and $F^{G}_{GGG}$ thus yields exact expressions for the generators $b_{1,G}$ and $b_{2,G}$. 
As we will discuss further in Section \ref{sec:R} however, the single exchange $R$ is not directly represented as a physical operation within the formalism of the quantum double model. Instead, the observable that naturally appears from commutation relations of ribbon operators is the monodromy (full braiding) operator $R^{2}$ \cite{buerschaper2009mapping, bombin2008family}.
In this protocol we therefore consider the full braiding operators
\begin{align}
    & B_{1,G}\equiv b_{1,G}^{2}= (R^{GG})^{2}, \nonumber \\ 
    & B_{2,G}\equiv b_{2,G}^{2}=F^{G \hspace{0.1cm} -1}_{GGG}(R^{GG})^{2}F^{G}_{GGG},
\end{align} describing complete evolutions of $G_{1}$ around $G_{2}$ and $G_{2}$ around $G_{3}$ respectively (see Figure \ref{fig:b1b2}).
Explicitly, $B_{2,G}$ has the following action on the anyonic basis $\{\ket{A},\ket{B},\ket{G}\}$
\begin{equation}
    B_{2,G} = \begin{pmatrix}
        \cos{\left(\frac{2\pi}{3}\right)} & -i\sin{\left(\frac{2\pi}{3}\right)} & 0\\
        -i\sin{\left(\frac{2\pi}{3}\right)} & \cos{\left(\frac{2\pi}{3}\right)} & 0 \\
        0 & 0 & \bar{\omega}
    \end{pmatrix}.
\end{equation}
Revealing a splitting in the qutrit subspace as this operation preserves the two-dimensional subspace $\text{span}(\ket{A},\ket{B})$ and its orthogonal complement $\text{span}(\ket{G})$. $B_{2,G}$ is thus a well-defined operation on a qubit encoded in terms of the reduced basis $\{\ket{A},\ket{B}\}$. Notably, inspection of this logical operation reveals that it does not belong to the single-qubit Clifford group $\mathcal{C}_{1}$.
Furthermore, creating two pairs of $G$ anyons from the vacuum and performing the braiding operation $B_{2,G}$ enables the preparation of the state
\begin{equation}
    \ket{\psi} = B_{2,G}\ket{A} = \cos{\left(\frac{2\pi}{3}\right)}\ket{A} -i\sin{\left(\frac{2\pi}{3}\right)}\ket{B},
\end{equation}
a magic state with $M_{2}(\ket{\psi})=\log{\left(\frac{16}{13}\right)}$. This non-stabilizer state constitutes a crucial resource for quantum computation, whereby the encoded non-Clifford action
enables the preparation of arbitrary quantum states using circuits composed solely of classically simulable Clifford operations \cite{laubscher2019universal}.

To illustrate the relative computational power of these $G$ anyons, we provide a brief comparison with an alternative non-Abelian subgroup of $\mathbf{D}(\mathbf{S}_{3})$, the charge subgroup $\{A,B,C\}$. 
As outlined in \cite{goel2023unveiling}, the non-Abelian chargeon $C$ has identical fusion rules to that of the $G$ anyon, such that a set of four $C$ anyons similarly encodes a logical qutrit. The form of the braiding matrix  
\begin{equation}
     R^{CC} = \begin{pmatrix}
        1 & 0 & 0 \\
        0 & -1 & 0 \\
        0 & 0 & 1
    \end{pmatrix},
\end{equation}
acting on the basis $\{\ket{A},\ket{B},\ket{C}\}$ however, yields $(R^{CC})^{2}=\mathbf{1}_{3}$. As a result the braiding operators $B_{1,C}=(R^{CC})^{2}=\mathbf{1}_{3}$ and $B_{2,C}=F^{C \hspace{0.1cm} -1}_{CCC}(R^{CC})^{2}F^{C}_{CCC}=\mathbf{1}_{3}$ act trivially on this fusion space, rendering this alternative subgroup unable to manifest non-trivial braiding statistics within the quantum double model.

This discussion thus serves to illustrate the potential computational power arising from the pairwise braiding of non-Abelian $G$ anyons.  
With the basic components of the $D(S_{3})$ quantum double model introduced in the next section, we thus seek to explicitly demonstrate the non-Clifford action of these braiding operations through simple operations on a lattice of $d=6$ qudits.

\section{Commutativity of Vertex and Plaquette Operators}
\label{sec:comm}

\begin{figure}[!t]
    \centering
    \includegraphics[width=0.45\textwidth]{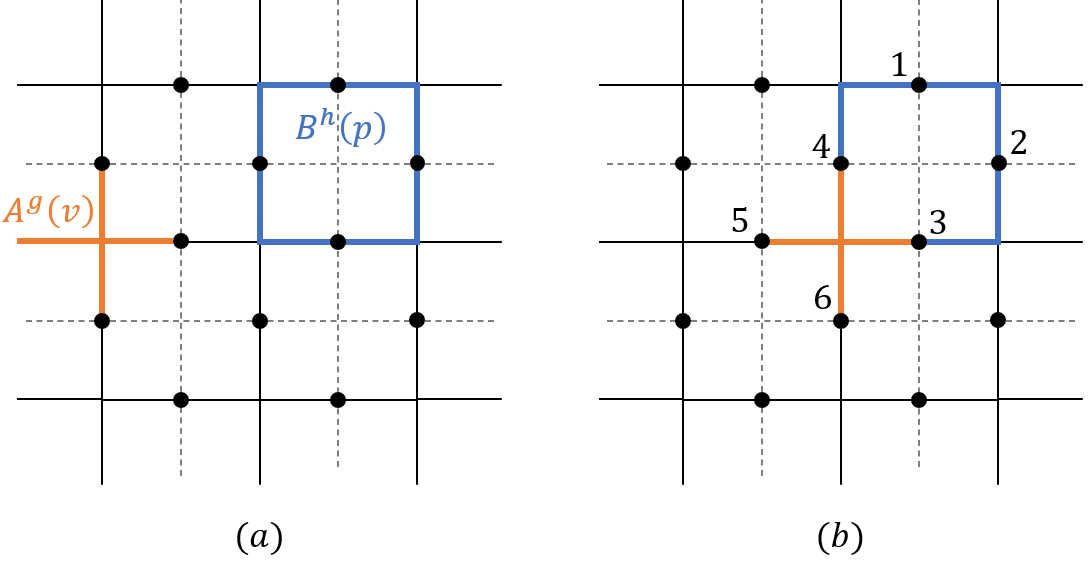}
    \caption{(a) A four plaquette lattice with a vertex operator $A^g(v)$ and plaquette operator $B^h(p)$ acting on different qudits. (b) The same plaquette but with vertex and plaquette operators overlapping on two qudits 3 and 4.}
    \label{fig:AB_commute}
\end{figure}

First, consider a plaquette operator $B^{e}(p)$ and a vertex operator $A(v)$. In calculating the commutation relation between any two such operators, there are two distinct cases as shown in Figure \ref{fig:AB_commute}. In the first case, when both $A(v)$ and $B^{e}(p)$ act on two different sets of four qudits as in Fig.\ref{fig:AB_commute}(a), it is clear that they will necessarily commute. In the alternative non-trivial case where $A(v)$ and $B^{e}(p)$ ‘overlap' and have non-trivial action on two of the same qudits on the lattice more care must be taken to demonstrate that these operators will commute. 
    
Consider, for example, the operators $A(v)$ and $B^{e}(p)$ acting on the six qudits shown in Figure \ref{fig:AB_commute}. For the labeling as shown these operators have the following forms
\begin{equation}
\begin{split}
    A(v) = \frac{1}{6}\sum_{g\in\mathbf{G}} \mathbf{1}\otimes\mathbf{1}\otimes L^{g}_{+}\otimes L^{g}_{+}\otimes L^{g}_{-}\otimes L^{g}_{-}, \\
    B^{e}(p) = \sum_{h_{1}h_{2}h_{3}h_{4}=e} T^{h_{1}}_{+} \otimes T^{h_{2}}_{-} \otimes T^{h_{3}}_{-}\otimes T^{h_{4}}_{+}\otimes \mathbf{1}\otimes \mathbf{1} .
\end{split} 
\end{equation} 
The action on all unlabelled qudits on this lattice is trivial and can thus be ignored.

In order to show that $[A(v),B^{e}(p)]\equiv A(v)B^{e}(p)-B^{e}(p)A(v)=0$ we expand out the form of each of the products,
\begin{equation}
\begin{split}
    & A(v)B^{e}(p) = \\ 
    & \frac{1}{6} \sum_{g\in\mathbf{G}}\sum_{h_{1}h_{2}h_{3}h_{4}=e}
    T^{h_{1}}_{+} \otimes T^{h_{2}}_{-} \otimes L^{g}_{+}T^{h_{3}}_{-}\otimes L^{g}_{+}T^{h_{4}}_{+}\otimes L^{g}_{-}\otimes L^{g}_{-}
    \label{eq:AB1}
\end{split}
\end{equation}
and
\begin{equation}
\begin{split}
    & B^{e}(p)A(v) = \\
    & \frac{1}{6} \sum_{g\in\mathbf{G}}\sum_{h_{1}h_{2}h_{3}h_{4}=e} T^{h_{1}}_{+} \otimes T^{h_{2}}_{-} \otimes T^{h_{3}}_{-}L^{g}_{+}\otimes T^{h_{4}}_{+}L^{g}_{+}\otimes L^{g}_{-}\otimes L^{g}_{-}.
    \label{eq:BA}
\end{split}
\end{equation}

In order to show that these two operators commute we refer to the commutation relations between the foundational $L^{g}_{\pm}$ and $T^{h}_{\pm}$ operators
\begin{equation}
    L^{g}_{+}T^{h}_{+}=T^{gh}_{+}L^{g}_{+}, \hspace{0.5cm} L^{g}_{-}T^{h}_{+}=T^{hg^{-1}}_{+}L^{g}_{-},
    \label{eq:com1}
\end{equation}
\begin{equation}
    L^{g}_{+}T^{h}_{-}=T^{hg^{-1}}_{+}L^{g}_{+}, \hspace{0.5cm} L^{g}_{-}T^{h}_{-}=T^{gh}_{-}L^{g}_{-},
    \label{eq:com2}
\end{equation}
as can be reproduced from the definitions of the action of $L^{g}_{\pm}$ and $T^{h}_{\pm}$ in Section \ref{sec:LatticeRep}.
The commutation identities in \eqref{eq:com1} and \eqref{eq:com2} yield $L^{g}_{+}T^{h_{3}}_{-}=T^{h_{3}g^{-1}}_{+}L^{g}_{+}$ and $L^{g}_{+}T^{h_{4}}_{+}=T^{gh_{4}}_{+}L^{g}_{+}$ such that \eqref{eq:AB1} becomes
\begin{equation}
\begin{split}
    A(v)B^{e}(p) = \frac{1}{6} \sum_{g\in\mathbf{G}}\sum_{h_{1}h_{2}h_{3}h_{4}=e} & T^{h_{1}}_{+} \otimes T^{h_{2}}_{-} \otimes T^{h_{3}g^{-1}}_{+}L^{g}_{+} \\
    & \otimes T^{gh_{4}}_{+}L^{g}_{+}\otimes L^{g}_{-}\otimes L^{g}_{-}.
\end{split}
\end{equation}
Introducing the relabelling $h_{1}^{\prime}=h_{1}$, $h_{2}^{\prime}=h_{2}$, $h_{3}^{\prime}=h_{3}g^{-1}$ and $h_{4}^{\prime}=g h_{4}$, we note that this set of new variables retains the initial condition $h_{1}^{\prime}h_{2}^{\prime}h_{3}^{\prime}h_{4}^{\prime}=h_{1}h_{2}h_{3}g^{-1}gh_{4}=h_{1}h_{2}h_{3}h_{4}=e$ such that
\begin{equation}
\begin{split}
    & A(v)B^{e}(p) = \\
    & \frac{1}{6} \sum_{g\in\mathbf{G}}\sum_{h_{1}^{\prime}h_{2}^{\prime}h_{3}^{\prime}h_{4}^{\prime}=e} T^{h_{1}^{\prime}}_{+} \otimes T^{h_{2}^{\prime}}_{-} \otimes T^{h_{3}^{\prime}}_{+}L^{g}_{+}\otimes T^{h_{4}^{\prime}}_{+}L^{g}_{+}\otimes L^{g}_{-}\otimes L^{g}_{-}.
    \label{eq:AB}
\end{split}
\end{equation}
In this way each term in \eqref{eq:AB} exactly matches one in \eqref{eq:BA} and the commutativity condition $[A(v),B^{e}(p)]=0$ is fuflfilled.
Crucially, for this and all other pairs of overlapping vertex and plaquette operators, this commutativity is ensured by choosing an orientation convention for the $L^{g}_{\pm}$ and $T^{h}_{\pm}$ operators such that one pair of $L^{g}_{\pm}$ and $T^{h}_{\pm}$ acting on the same qudit have the matching parity and the other have opposite parity. This ensures the condition $h_{1}^{\prime}h_{2}^{\prime}h_{3}^{\prime}h_{4}^{\prime}=e$ such that the two operators commute.

Similarly, for overlapping vertex (plaquette) operators, the $L^{g}_{\pm}$ ($T^{h}_{\pm}$) operators acting on the single qudit point of overlap  necessarily have opposite parity. For example, for the general case of two neighbouring vertex operators $A(v_{i})$ and $A(v_{j})$
    \begin{equation}
        [A(v_{i}),A(v_{j})] \propto \sum_{g} \sum_{h} [L^{g}_{\pm},L^{h}_{\mp}] = 0.
    \end{equation}

\section{$R$ matrix from ribbon operators}
\label{sec:R_full}

In Section \ref{sec:Rprod} we have shown how the exchange matrix $(R^{GG})^2$, describing the braiding of two $G$ anyons, may be reproduced with projective measurements onto states constructed by exchanging the order of the ribbon operators $F^{G}_{\rho_{1}}$ and $F^{G}_{\rho_{2}}$, replicating Equation \ref{eq:Rbraid}. This result may also be verified with direct evaluation of the operator products $F^{G}_{\rho_{1}}F^{G}_{\rho_{2}}$ and $F^{G}_{\rho_{2}}F^{G}_{\rho_{1}}$ in the basis $\{e,c,c^{2},t,tc,tc^{2}\}$. By comparing the form of each operator, the $R$ matrix can then be determined in the anyonic basis $\{A,B,G\}$. Such an approach can be experimentally relevant when quantum operator tomography is available.

By representing the ribbons operators $F^G_{\rho_1}$ and $F^G_{\rho_2}$ fully in terms of their elements of the isolated two lattice sites they cross over, we can find all the elements of the crossed terms
\begin{widetext}
\begin{equation}
    \begin{split}
        F^G_{\rho_1}F^G_{\rho_2} &= \left((T^{e}_{-} + \omega T^{c}_{-} + \bar{\omega} T^{c^{2}}_{-}) \otimes L^{c}_{-} + (T^{e}_{-} + \bar{\omega} T^{c}_{-} + \omega T^{c^{2}}_{-}) \otimes L^{c^{2}}_{-} \right) \\
        &\qquad\qquad\times \left(L^{c}_{+}\otimes (T^{e}_{-} + \omega T^{c}_{-} + \bar{\omega} T^{c^{2}}_{-}) + L^{c^{2}}_+\otimes (T^{e}_{-} + \bar{\omega} T^{c}_{-} + \omega T^{c^{2}}_{-}) \right), \\
        &= \left(\sum_h \left(\ket{e}\bra{e} + \Bar{\omega}\ket{c}\bra{c} + \omega\ket{c^2}\bra{c^2}\right)\otimes \ket{c^2h}\bra{h} + \left(\ket{e}\bra{e} + \omega\ket{c}\bra{c} + \Bar{\omega}\ket{c^2}\bra{c^2}\right)\otimes \ket{ch}\bra{h} \right)\\
        &\times \left(\sum_g \ket{gc}\bra{g}\otimes \left(\ket{e}\bra{e} + \Bar{\omega}\ket{c}\bra{c} + \omega\ket{c^2}\bra{c^2}\right) + \ket{gc^2}\bra{g}\otimes \left(\ket{e}\bra{e} + \omega\ket{c}\bra{c} + \Bar{\omega}\ket{c^2}\bra{c^2}\right) \right), \\
        &= \ket{e,c^{2}}\bra{c,e} + \ket{e,c^{2}}\bra{c^{2},e} + \ket{e,c}\bra{c,e} + \ket{e,c}\bra{c^{2},e} + \omega\ket{c^{2},c^{2}}\bra{e,e}+ \omega\ket{c^{2},c^{2}}\bra{c,e}\\ 
        &+ \bar{\omega}\ket{c^{2},c}\bra{e,e} + \bar{\omega}\ket{c^{2},c}\bra{c,e} +\bar{\omega}\ket{c,c^{2}}\bra{e,e} + \bar{\omega}\ket{c,c^{2}}\bra{c^{2},e} +\omega\ket{c,c}\bra{e,e} + \omega\ket{c,c}\bra{c^{2},e} \\
        &+ \omega\ket{e,e}\bra{c,c} 
        +\bar{\omega}\ket{e,e}\bra{c^{2},c} + \omega\ket{e,c^{2}}\bra{c,c} + \bar{\omega}\ket{e,c^{2}}\bra{c^{2},c} + \bar{\omega}\ket{c^{2},e}\bra{e,c} +\ket{c^{2},e}\bra{c,c} \\ 
        &+ \ket{c^{2},c^{2}}\bra{e,c} + \omega\ket{c^{2},c^{2}}\bra{c,c} 
        + \omega\ket{c,e}\bra{e,c} + \ket{c,e}\bra{c^{2},c} + \ket{c,c^{2}}\bra{e,c} + \bar{\omega}\ket{c,c^{2}}\bra{c^{2},c} \\
        &+ \bar{\omega}\ket{e,c}\bra{c,c^{2}} + \omega\ket{e,c}\bra{c^{2},c^{2}} + \bar{\omega}\ket{e,e}\bra{c,c^{2}} 
        + \omega\ket{e,e}\bra{c^{2},c^{2}} + \ket{c^{2},c}\bra{e,c^{2}} +\bar{\omega}\ket{c^{2},c}\bra{c,c^{2}} \\
        &+ \omega\ket{c^{2},e}\bra{e,c^{2}} + \ket{c^{2},e}\bra{c,c^{2}} + \ket{c,c}\bra{e,c^{2}} + \omega\ket{c,c}\bra{c^{2},c^{2}} 
        + \bar{\omega}\ket{c,e}\bra{e,c^{2}} + \ket{c,e}\bra{c^{2},c^{2}},
    \end{split}
\end{equation}
\begin{equation}
    \begin{split}
        F^{G}_{\rho_{2}}F^{G}_{\rho_{1}} &= \omega\ket{e,c^{2}}\bra{c,e} + \bar{\omega}\ket{e,c^{2}}\bra{c^{2},e} + \bar{\omega}\ket{e,c}\bra{c,e} + \omega\ket{e,c}\bra{c^{2},e} + \bar{\omega}\ket{c^{2},c^{2}}\bra{e,e}+ \ket{c^{2},c^{2}}\bra{c,e}\\ 
        &+ \omega\ket{c^{2},c}\bra{e,e} + \ket{c^{2},c}\bra{c,e} +\omega\ket{c,c^{2}}\bra{e,e} + \ket{c,c^{2}}\bra{c^{2},e} +\bar{\omega}\ket{c,c}\bra{e,e} + \ket{c,c}\bra{c^{2},e} \\
        &+ \bar{\omega}\ket{e,e}\bra{c,c} 
        +\omega\ket{e,e}\bra{c^{2},c} + \ket{e,c^{2}}\bra{c,c} + \ket{e,c^{2}}\bra{c^{2},c} + \ket{c^{2},e}\bra{e,c} +\bar{\omega}\ket{c^{2},e}\bra{c,c} \\ 
        &+ \bar{\omega}\ket{c^{2},c^{2}}\bra{e,c} + \bar{\omega}\ket{c^{2},c^{2}}\bra{c,c} 
        + \ket{c,e}\bra{e,c} + \omega\ket{c,e}\bra{c^{2},c} + \omega\ket{c,c^{2}}\bra{e,c} + \omega\ket{c,c^{2}}\bra{c^{2},c} \\
        &+ \ket{e,c}\bra{c,c^{2}} + \ket{e,c}\bra{c^{2},c^{2}} + \ket{e,e}\bra{c,c^{2}} + \bar{\omega}\ket{e,e}\bra{c^{2},c^{2}} + \omega\ket{c^{2},c}\bra{e,c^{2}} +\omega\ket{c^{2},c}\bra{c,c^{2}} \\
        &+ \ket{c^{2},e}\bra{e,c^{2}} + \omega\ket{c^{2},e}\bra{c,c^{2}} + \bar{\omega}\ket{c,c}\bra{e,c^{2}} + \bar{\omega}\ket{c,c}\bra{c^{2},c^{2}} 
        + \ket{c,e}\bra{e,c^{2}} + \bar{\omega}\ket{c,e}\bra{c^{2},c^{2}}.
    \end{split}
\end{equation}
\end{widetext}
Comparing these two expressions, we find that the elements of $F^{G}_{\rho_{2}}F^{G}_{\rho_{1}}$ gain an additional phase relative to those of $F^{G}_{\rho_{1}}F^{G}_{\rho_{2}}$ according to the following prescription
\begin{equation}
    \ket{g_{1}, g_{2}}\bra{h_{1}, h_{2}} \rightarrow \begin{cases}
      \bar{\omega} \ket{g_{1}, g_{2}}\bra{h_{1}, h_{2}} & \text{if } g_{1}g_{2}=h_{1}h_{2}\\
      \omega \ket{g_{1}, g_{2}}\bra{h_{1}, h_{2}} & \text{otherwise}.
    \end{cases} 
    \label{eq:g1g2h1h2}
\end{equation}

To re-interpret this result into the anyonic basis as in \eqref{eq:R_matrix}, we consider the flux these ribbon operator products create on the shared plaquette, $p$. As the vacuum state is stabilised by the projector $B^{e}(p)$ it has trivial flux. As a result it consists of a superposition over all group elements for which $h_{1}h_{2}h_{3}h_{4}=e$ on that plaquette. Under the action of $\ket{g_{1},g_{2}}\bra{h_{1},h_{2}}$, if $g_{1}g_{2}=h_{1}h_{2}$, then the condition $g_{1}g_{2}h_{3}h_{4}=e$ is preserved. This condition corresponds to the fusion outcomes $A$ and $B$ that have trivial flux, that gain a factor $\bar{\omega}$, as seen from \eqref{eq:g1g2h1h2}. When $g_{1}g_{2}\neq h_{1}h_{2}$, then $g_{1}g_{2}h_{3}h_{4}\neq e$ and the two ribbons have fused to produce a pair of $G$ anyons with non-trivial flux. As seen from \eqref{eq:g1g2h1h2}, this fusion outcome acquires a phase factor $\omega$ as in the final diagonal element of \eqref{eq:R_matrix}.

\section{Analytics for $F$ matrix}
\label{sec:Folap}

We will now present the main calculations for the amplitudes that determine the elements of the $F$ matrix squared. By taking the inner products of our ribbon fusion states as in \eqref{eq:FRphase}, we find that for each combination $i,j=A,B,G$ a phase factor may be extracted,
\begin{equation}
    \begin{split}
        &  \braket{\psi_2(j) | \psi_1(i)} \\
        & = \bra{\zeta} \left(A^{G}(v) F^{G}_{\rho_{1}} A^{j}(v) F^{G}_{\rho_{2}} F^{G}_{\rho_{2}}\right)^\dagger A^{G}(v) F^{G}_{\rho_{2}} A^{i}(v) F^{G}_{\rho_{2}} F^{G}_{\rho_{1}} \ket{\zeta}, \\
        & = \bra{\zeta} (F^{G}_{\rho_{2}})^{2} A^{j}(v) F^{G}_{\rho_{1}} A^{G}(v) F^{G}_{\rho_{2}} A^{i}(v) F^{G}_{\rho_{2}} F^{G}_{\rho_{1}} \ket{\zeta}, \\
       & = \bra{\zeta} F^{j}_{\rho_{2}} F^{G}_{\rho_{1}} A^{G}(v) A^{G}(v) F^{G}_{\rho_{2}} A^{i}(v) F^{G}_{\rho_{2}} F^{G}_{\rho_{1}} \ket{\zeta}, \\
        & = (R^{Gj}_G)^2 \bra{\zeta} F^{G}_{\rho_{1}} F^{j}_{\rho_{2}} A^{G}(v) F^{G}_{\rho_{2}} A^{i}(v) F^{G}_{\rho_{2}} F^{G}_{\rho_{1}} \ket{\zeta},
    \end{split}
\end{equation}
where we have used the Hermiticity of each of our operators, $A^{j}(v)(F^{G}_{\rho_{2}})^{2}\ket{\zeta}=A^{j}(v)(F^{A}_{\tau_{2}}+F^{B}_{\tau_{2}}+F^{G}_{\rho_{2}})\ket{\zeta}=F^{j}_{\tau/\rho_{2}}\ket{\zeta}$ and $A^{G}(v)F^{G}_{\rho_{1}}F^{j}_{\tau/\rho_{2}}\ket{\zeta}={\overline{(R^{Gj}_{G}})^2}A^{G}(v)F^{j}_{\tau/\rho_{2}}F^{G}_{\rho_{1}}\ket{\zeta}$. The string operator $F^j_{\tau_2}$ with $j=A$ or $B$ is the direct triangle in the path $\rho_2$, producing a pair of $A$ or $B$ anyons as in \eqref{eq:FA} and $\eqref{eq:FB}$ respectively. 

All these possible inner products can be reduced to phases and inner amplitudes with manipulations of the ribbon operators. We see that if $j = A$ or $B$ and $i=A,B$ or $G$, we can take
\begin{equation}
    \begin{split}
        & \braket{\psi_2(j) | \psi_1(i)} \\
        & = (R^{Gj}_G)^2 \bra{\zeta} F^{G}_{\rho_{1}} F^{j}_{\rho_{2}} A^{G}(v) F^{G}_{\rho_{2}} A^{i}(v) F^{G}_{\rho_{2}} F^{G}_{\rho_{1}} \ket{\zeta}, \\
        & = (R^{Gj}_G)^2\bra{\zeta} F^{G}_{\rho_{1}} F^{G}_{\rho_{2}} A^{i}(v) A^{i}(v) F^{G}_{\rho_{2}} F^{G}_{\rho_{1}} \ket{\zeta}, \\
        & = (R^{Gj}_G)^2\frac{1}{N_{i}^{2}} \braket{\phi_{21}(i) | \phi_{21}(i)}, \\
        &= \frac{(R^{Gj}_G)^2}{N_{i}^{2}},
    \end{split}
\end{equation}
for the normalised states $\ket{\phi_{21}(i)}$ as defined in \eqref{eq:1221}.

When $j=G$ and $i=A$ or $B$ we have
\begin{equation}
    \begin{split}
        & \braket{\psi_2(G) | \psi_1(i)} \\
        & = (R^{GG}_G)^2 \bra{\zeta} F^{G}_{\rho_{1}} F^{G}_{\rho_{2}} A^{G}(v) F^{G}_{\rho_{2}} A^{i}(v) F^{G}_{\rho_{2}} F^{G}_{\rho_{1}} \ket{\zeta}, \\
        & = (R^{GG}_G)^2 \bra{\zeta} F^{G}_{\rho_{1}} (F^{G}_{\rho_{2}})^2 A^{i}(v) F^{G}_{\rho_{2}} F^{G}_{\rho_{1}} \ket{\zeta}, \\
        & = (R^{GG}_G)^2 \bra{\zeta} F^{G}_{\rho_{1}} F^{G}_{\rho_{2}} A^{i}(v) A^{i}(v) F^{G}_{\rho_{2}} F^{G}_{\rho_{1}} \ket{\zeta}, \\
        & = \frac{(R^{GG}_G)^2}{N_{i}^{2}}.
    \end{split}
\end{equation}
And finally, for the case $i=j=G$
\begin{equation}
    \begin{split}
        & \braket{\psi_2(G) | \psi_1(G)} \\
        & = (R^{GG}_G)^2 \bra{\zeta} F^{G}_{\rho_{1}} F^{G}_{\rho_{2}} A^{G}(v) F^{G}_{\rho_{2}} A^{G}(v) F^{G}_{\rho_{2}} F^{G}_{\rho_{1}} \ket{\zeta}, \\
        & = \frac{(R^{GG}_G)^2}{N_{G}^{2}} \bra{\phi_{21}(G)} F^{G}_{\rho_{2}} \ket{\phi_{21}(G)}.
    \end{split}
\end{equation}
The operator $F^{G}_{\rho_{2}}$ as defined in Eq.~\eqref{eq:rho_2} is constructed from terms of the form $L^{g}_{+}\otimes T^{h}_{-}$. The non-zero matrix elements of the operators $T^{h}_{-}$ lie on the main diagonal, whereas $L^{g}_{+}$ act as permutation operators between group elements such that the resultant operator $F^{G}_{\rho_{2}}$ is traceless and we obtain the required $\braket{\psi_2(G) | \psi_1(G)} = 0$.

These analytical values for each of the overlaps $\braket{\psi_2(j) | \psi_1(i)}$ may be substituted into Equation \eqref{eq:Ffull} in order to determine each of the values of the $F$ matrix squared.

\section{Derivation of Two Qudit Dense Encoding}
\label{sec:2qudits}

The derivation of $R^{GG}$ and $\mathcal{F}^{G}_{GGG}$ from the minimal physical system may be represented as the calculation of overlaps of the form
\begin{equation}
    \bra{\eta} O_{a,4}^{\dagger}O_{b,4}\ket{\eta},
\end{equation}
where $O_{i,4}$ indicates a matrix acting on the $6^{4}$-dimensional Hilbert space of a single plaquette and $\ket{\eta}$ is the ground state of the Kitaev Hamiltonian on this system
\begin{equation}
\mathcal{H}_{4} = -B(p) - \sum_{i=1}^{4} A(v_{i}),
\label{eq:H4_app}
\end{equation}
where the vertex operators, $A(v_{i})$, act non-trivially on only two qudits, such that for example
\begin{equation}
    A(v_{3}) = \frac{1}{6} \sum_{g\in\mathbf{G}} \mathbf{1}_{6}\otimes \mathbf{1}_{6}\otimes L^{g}_{+}\otimes L^{g}_{-}.
    \label{eqn:pro}
\end{equation}
We have also shown, however, that each operator acting on $\ket{\eta}$ in the states \eqref{eq:psi_states2} and \eqref{eq:psi_states3}, only acts non-trivially on qudits 3 and 4. We may therefore rewrite
\begin{align}
    \bra{\eta}O_{a,4}^{\dagger}O_{b,4}\ket{\eta} &= \bra{\eta} (\mathbf{1}_{36}\otimes O_{a}^{\dagger})(\mathbf{1}_{36}\otimes O_{b})\ket{\eta} \\
    &= \bra{\eta} (\mathbf{1}_{36}\otimes O_{a}^{\dagger}O_{b})\ket{\eta} \\ 
    &= \bra{\eta} \begin{pmatrix}
        O_{a}^{\dagger}O_{b} & 0 & \dots & 0 \\
        0 & O_{a}^{\dagger}O_{b} & \dots & 0 \\
        \vdots & \vdots & \ddots & \vdots \\
        0 & 0 & \dots & O_{a}^{\dagger}O_{b}
    \end{pmatrix} \ket{\eta}.
\end{align}
Each element `$O_{a}^{\dagger}O_{b}$' is itself a $36\times 36$ matrix, such that this expression may be further simplified by decomposing $\ket{\eta}$ into a set of $36$-dimensional vectors.
It is found that $\ket{\eta}$ may be written in the following form
\begin{equation}
    \ket{\eta} = \begin{pmatrix}
        \ket{\psi_{1}} \\
        \ket{\psi_{2}} \\
        \ket{\psi_{3}} \\
        \ket{\psi_{4}} \\
        \ket{\psi_{5}} \\
        \ket{\psi_{6}}
    \end{pmatrix} = \begin{pmatrix}
        \ket{\psi_{1}} \\
        (L_{-}^{c}\otimes\mathbf{1}_{36})\ket{\psi_{1}} \\
        (L_{-}^{c^{2}}\otimes\mathbf{1}_{36})\ket{\psi_{1}} \\
        (L_{-}^{t}\otimes\mathbf{1}_{36})\ket{\psi_{1}} \\
        (L_{-}^{tc}\otimes\mathbf{1}_{36})\ket{\psi_{1}} \\
        (L_{-}^{tc^{2}}\otimes\mathbf{1}_{36})\ket{\psi_{1}}
    \end{pmatrix},
\end{equation}
where $\ket{\psi_{1}}$ is a $6^{3}$-dimensional vector of the form
\begin{equation}
    \ket{\psi_{1}} = \begin{pmatrix}
        \ket{\psi_{e}} \\
        \ket{\psi_{c}} \\
        \ket{\psi_{c^{2}}} \\
        \ket{\psi_{t}} \\
        \ket{\psi_{tc}} \\
        \ket{\psi_{tc^{2}}}
    \end{pmatrix}, \hspace{0.2cm} \ket{\psi_{g}} = \frac{1}{\sqrt{216}}\sum_{g_{1}g_{2}=g} \ket{g_{1},g_{2}}.
\end{equation}
These $\ket{\psi_{g}}$ are the $36$-dimensional vectors on which $O_{a}^{\dagger}O_{b}$ will act.

The single-qudit right multiplication operators $L_{-}^{g}$ are essentially permutation operators such that, for example
\begin{equation}
    \ket{\psi_{2}} = (L^{c}_{-}\otimes\mathbf{1}_{36})\ket{\psi_{1}} = \begin{pmatrix}
        \ket{\psi_{c}} \\
        \ket{\psi_{c^{2}}} \\
        \ket{\psi_{e}} \\
        \ket{\psi_{tc}} \\
        \ket{\psi_{tc^{2}}} \\
        \ket{\psi_{t}}
    \end{pmatrix}.
\end{equation}

We therefore see that each 36-dimensional vector $\ket{\psi_{g}}$ will appear in each of $\ket{\psi_{1}},\ket{\psi_{2}},\dots,\ket{\psi_{6}}$ once and therefore in total in $\ket{\eta}$ six times.
Hence, all ground state expectation values of operators acting only non-trivially on qudits 3 and 4 may be decomposed as the sum of six inner products within this reduced system
\begin{equation}
    \bra{\eta} (\mathbf{1}_{36}\otimes O_{a}^{\dagger}O_{b})\ket{\eta} = 6\sum_{g\in G} \bra{\psi_{g}}O_{a}^{\dagger}O_{b}\ket{\psi_{g}}.
\end{equation}
In this way, the $R$ and $F$ matrices may be reconstructed with the consideration of the expectation values of the operators in Equations \eqref{eq:psi_states2} and \eqref{eq:psi_states3} with respect to the set of two-qudit states $\{\ket{\psi_{g}}\}$.




\end{appendix}

\end{document}